\documentclass[12pt,preprint]{aastex}

\usepackage{rotating}

\shorttitle{X-Ray Probing of Young Ejecta-Dominated SNRs}  
\shortauthors{Katsuda et al.}

\begin{document}

\title{Kepler's Supernova: An Overluminous Type~Ia Event Interacting with a Massive Circumstellar Medium at a Very Late Phase}

\author{Satoru Katsuda\altaffilmark{1}, Koji Mori\altaffilmark{2}, 
Keiichi Maeda\altaffilmark{3, 4}, Masaomi Tanaka\altaffilmark{5}, 
Katsuji Koyama\altaffilmark{6, 7}, Hiroshi Tsunemi\altaffilmark{6}, 
Hiroshi Nakajima\altaffilmark{6}, Yoshitomo Maeda\altaffilmark{1}, 
Masanobu Ozaki\altaffilmark{1}, \&\\ 
Robert Petre\altaffilmark{8}
}

\altaffiltext{1}{Institute of Space and Astronautical Science (ISAS), Japan Aerospace Exploration Agency (JAXA), 3-1-1 Yoshinodai, Chuo, Sagamihara, Kanagawa 252-5210, Japan}

\altaffiltext{2}{Department of Applied Physics and Electronic Engineering, Faculty of Engineering, University of Miyazaki, 1-1 Gakuen Kibanadai-Nishi, Miyazaki 889-2192}

\altaffiltext{3}{Department of Astronomy, Kyoto University, Kitashirakawa-Oiwake-cho, Sakyo-ku, Kyoto 606-8502, Japan}

\altaffiltext{4}{Kavli Institute for the Physics and Mathematics of the Universe (WPI), University of Tokyo, 5-1-5 Kashiwanoha, Kashiwa, Chiba 277-8583, Japan}

\altaffiltext{5}{National Astronomical Observatory of Japan, Mitaka, Tokyo 181-8588, Japan}

\altaffiltext{6}{Department of Earth and Space Science, Osaka University, 1-1 Machikaneyama-cho, Toyonaka, Osaka 560-0043}

\altaffiltext{7}{Department of Physics, Graduate School of Science, Kyoto University, Kitashirakawa Oiwake-cho, Sakyo-ku, Kyoto 606-8502}

\altaffiltext{8}{Astrophysics Science Division, NASA Goddard Space Flight Center, Greenbelt, MD 2077, USA}

\begin{abstract}

We have analyzed {\it XMM-Newton}, {\it Chandra}, and {\it Suzaku} observations of Kepler's supernova remnant (SNR) to investigate the properties of both the SN ejecta and the circumstellar medium (CSM).  For comparison, we have also analyzed two similarly-aged, ejecta-dominated SNRs: Tycho's SNR, thought to be the remnant of a typical Type Ia SN, and SNR~0509-67.5 in the Large Magellanic Cloud, thought to be the remnant of an overluminous (SN1991T-like) Type Ia SN.  By simply comparing the X-ray spectra, we find that line intensity ratios of iron-group elements (IGE) to intermediate-mass elements (IME) for Kepler's SNR and SNR~0509-67.5 are much higher than those for Tycho's SNR.  We therefore argue that Kepler is the product of an overlumious Type Ia SN.  
This inference is supported by our spectral modeling, which reveals the IGE and IME masses respectively to be 0.95$^{+0.34}_{-0.37}$\,M$_\odot$ and 0.12$^{+0.19}_{-0.05}$\,M$_\odot$ (Kepler's SNR), 0.75$^{+0.51}_{-0.15}$\,M$_\odot$ and 0.34$^{+0.08}_{-0.25}$\,M$_\odot$ (SNR~0509-67.5), and 0.35$^{+0.55}_{-0.15}$\,M$_\odot$ and 0.70$^{+0.12}_{-0.28}$\,M$_\odot$ (Tycho's SNR).  
We find that the CSM component in Kepler's SNR consists of tenuous diffuse gas ($\sim$0.3\,M$_\odot$) present throughout the entire remnant, plus dense knots ($\sim$0.035\,M$_\odot$).  Both of these components have an elevated N abundance (N/H$\sim$4 times the solar value), suggesting that they originate from CNO-processed material from the progenitor system.  The mass of the diffuse CSM allows us to infer the pre-SN  mass-loss rate of the system to be $\sim1.5\times10^{-5}$($v_\mathrm{w}$/10\,km\,s$^{-1}$)\,M$_\odot$\,yr$^{-1}$, in general agreement with results from recent hydrodynamical simulations.  The dense knots have slow optical proper motions as well as relatively small X-ray--measured ionization timescales, which indicates that they were located a few pc away from the progenitor system and were only recently heated by forward shocks.  Therefore, we argue that Kepler's SN was an overluminous (91T-like) event that started to interact with massive CSM a few hundred years after the explosion.  This supports the possible link between 91T-like SNe and the so-called ``Ia-CSM" SNe --- a rare class of SNe Ia associated with massive CSM.  The link implies that $\sim$10\% of SNe Ia are associated with massive CSM which most likely originates from a companion star in a single degenerate progenitor system.

\end{abstract}

\section{Introduction}

Type Ia supernovae (SNe Ia) are important astrophysical objects, as they play a central role in measuring the accelerative expansion of the universe and are primary source of iron-group elements (IGE) in the universe.  However, the progenitor system(s) and detailed explosion mechanism(s) have been a matter of considerable debate \citep[e.g.,][for a recent review]{Maoz2014}.  It has been believed that SNe Ia are runaway thermonuclear explosions of a C/O white dwarf (WD), as it approaches the Chandrasekhar mass ($\sim$1.4\,M$_\odot$).  There are mainly two long-standing competing scenarios for how a WD increases its mass: (1) the single-degenerate model, in which a WD accretes material from a non-degenerate companion star in a close binary system \citep{Whelan1973}; and (2) the double-degenerate model, in which two C/O WDs merge \citep{Iben1984,Webbink1984}.  Therefore, identifying the progenitor stars (as a means of discriminating between the two scenarios) is an area of considerable activity in SN Ia studies.

There are many observational approaches to identifying progenitors, including studies of potential progenitor populations \citep{Hachisu2001}, searches for companions to the progenitor systems in optical pre/post-explosion images \citep[e.g.,][]{Li2011b,Foley2014,McCully2014} and surviving companions in nearby supernova remnants (SNRs) \citep[e.g.,][]{Schaefer2012,Bedin2014}, and evidence of interactions between the SN ejecta and the companion star \citep[e.g.,][]{Maeda2014}.  In addition, recent detection of a circumstellar medium in some SNe Ia (CSM: environmental gas modified by the mass loss from the progenitor systems) have opened a new window to infer progenitor systems.  

The presence of a H-rich CSM itself supports the SD scenario, since such a CSM is not generally expected around two WDs in the DD scenario.  Analysis of the CSM signature (both absorption and emission) offers further insight into the progenitor system.  For example, time-variable Na I D absorption features, which were found in the normal Type Ia SN~2006X and a few others \citep{Patat2007,Blondin2009,Simon2009}, could be interpreted to arise from expanding shells (or knots) due to successive nova eruptions of the progenitor \citep{Patat2007}.  Later, \citet{Sternberg2011} found that about half of all sampled SNe Ia (12/22) in their sample show blueshifted Na I D absorption lines, while only a quarter of them (5/22) show redshifted absorption, and the remaining quarter (5/22) show single or symmetric absorption profiles.  They suggested that the biased detection of the blueshifted systems means a large fraction of SNe Ia ($\gtrsim$25\%) are associated with progenitors' gas outflows (otherwise, we would see the same fractions of blueshifted and redshifted profiles).  From the absorption column densities ($N_\mathrm{Na I}\sim10^{12}$\,cm$^{-2}$), the masses of the CSM shells have been estimated to be $\lesssim$0.01\,M$_\odot$ \citep{Patat2007}, which are consistent with successive recurrent nova eruptions that swept up the stellar wind of the companion.  Therefore, these CSM properties can be best explained by a symbiotic recurrent nova progenitor, similar to RS Ophiuchi.

Evidence for much more massive CSM ($\sim$0.01--5\,M$_\odot$) has been found in several SNe Ia \citep[SN~2002ic, SN~2005gj, SN~1997cy, SN~1999E, SN~2008J, PTF11kx, and some other candidates:][references therein]{Hamuy2003,Silverman2013}.  These SNe are called ``SNe Ia-CSM" which are characterized by a narrow H$\alpha$ line on top of an overluminous (SN~1991T-like or 91T-like) Ia template in the optical spectra.  SNe Ia-CSM comprise only a small fraction of SNe Ia: 0.1--1\,\% \citep{Dilday2012}, albeit a true fraction may be somewhat larger, given a possibly significant amount of SNe Ia-CSM that are incorrectly classified as Type IIn because of their spectral resemblance \citep{Leloudas2013}.  While there are several proposed progenitor systems for Ia-CSM objects, none of them is yet conclusive.  

Kepler's SNR, the remains of SN~1604 A.D., is one of a handful of SNRs which are associated with historical Galactic SNe.  Located at only 3--7\,kpc \citep[][references therein]{Kerzendorf2014}, this remnant provides a precious opportunity to study details of SN Ia explosions and the progenitor system.  It belongs to a rare class of Type Ia SNRs \citep{Kinugasa1999,Reynolds2007} that exhibit emission from a CSM, evidenced by the extremely high ambient density ($\sim$100\,cm$^{-3}$) at $\gtrsim$500\,pc out of the Galactic plane and a N overabundance \citep{Blair1991,Gerardy2001}.  Optical monitoring of 50 long-lived CSM knots revealed that they are moving fairly slowly, with a space velocity of $\sim$280\,km\,s$^{-1}$ at a distance of 4.5\,kpc \citep{Bandiera1991}; reminiscent of quasi-stationary flocculi in Cassiopeia~A.  The trace-back time was measured to be $3.2\pm1.2\times10^4$\,yrs, which should be taken as a lower limit, given possible acceleration by the SN forward shock.

\citet{Douvion2001} found infrared (IR) emission from warm dust.  Since the IR intensity map is similar to the distribution of the dense optical CSM knots, the authors argued that the IR emission arises from dust in the shocked CSM.  \citet{Blair2007} further supported this picture based on {\it Spitzer Space Telescope} observations, and estimated the total shocked CSM mass to be $\sim$1\,M$_\odot$ from their relatively cursory analysis of X-ray data.  In addition to the CSM knots, the presence of a diffuse CSM has been inferred using followup {\it Spitzer} observations by \citet{Williams2012}.  Their work indicates that the CSM dust consists of two components: lower-temperature ($\sim$80--100\,K) dust behind fast shocks ($>$1000\,km\,s$^{-1}$) and penetrating into the tenuous ambient medium produces the majority of the IR emission; higher-temperature ($>$150\,K) dust behind slower shocks (a few hundred km\,s$^{-1}$) penetrating into high density material ($n\sim50-250$\,cm$^{-3}$) coincides with the optical dense knots.\footnote{We note that dust is mainly heated by collisions in shocked gases, and that the efficiency of collisional heating increases with the density of the shocked gas.}  The presence of a CSM is also supported from recent hydrodynamic simulations \citep{Velazquez2006,Chiotellis2012,Toledo-roy2014,Burkey2013}, where a high mass-loss rate of $\dot{M}\sim10^{-5}$\,M$_\odot$\,yr$^{-1}$, suggestive of dense winds from an asymptotic giant branch (AGB) star, can reasonably explain the observed expansion index and the north-south asymmetric morphology of Kepler's SNR.  

X-rays arise predominantly from SN ejecta, which are not detected at either optical or IR wavelengths.  The X-ray spectrum exhibits strong Fe L lines and weak O K lines. A multi-component non-equilibrium ionization modeling suggests a Type Ia SN rather than a core-collapse SN as its origin \citep[][references therein for a history of X-ray observations]{Kinugasa1999}.  From deep {\it Chandra} observations, \citet{Reynolds2007} confirmed the high Fe/O ratio and found no evidence for a neutron star, further supporting the Type Ia origin.  \citet{Patnaude2012} claimed that Kepler's SN was likely to be a 91T-like event that produced $\sim$1\,M$_\odot$ of $^{56}$Ni, by comparing the observed X-ray spectrum with simulated spectra based on their hydrodynamic modeling at 400\,yrs after the explosion.  More recently, \citet{Park2013} performed a deep {\it Suzaku} observation, and found a high Mn/Cr line intensity ratio of $\sim$0.6, leading them to suggest a supersolar metallicity ($\sim$3 times the solar value) of the progenitor star.  

Here, we investigate X-ray spectra of Kepler's SNR and two other young Ia SNRs, Tycho's SNR and SNR~0509-67.5, for comparison, by making the best use of available data from currently operational X-ray observatories.  In the following section 2, we summarize the observations used.  In sections 3 and 4, we perform X-ray spectral analyses, focusing mainly on the SN ejecta and the CSM, respectively.  Finally, we discuss the observational results and lead to conclusions.

\section{Observations and Data Reduction}

We use archival data obtained by {\it XMM-Newton}, {\it Chandra}, and {\it Suzaku}, as summarized in Table~\ref{tab:obs}.  While our object of interest is Kepler's SNR, we also analyze two similarly-aged Ia remnants, Tycho's SNR and SNR~0509-67.5 in the Large Magellanic Cloud (LMC).  Detailed classifications have already been determined for these two remnants, based on their light-echo spectra:  Tycho is thought to be a prototypical Type Ia \citep{Krause2008}, SNR~0509-67.5 is thought to be ``SN~1991T-like" \citep{Rest2005}.  Thus these two remnants can serve as ideal reference targets to study the amount of IGE in Kepler's SN.

We reprocessed the raw data, using version 13.5.0 of {\it XMM-Newton} Science Analysis Software, version 4.6 of {\it Chandra} Interactive Analysis of Observations, and version 22 of the {\it Suzaku} software, together with the latest versions of the calibration files at the analysis phase.  For the X-ray Imaging Spectrometer \citep[XIS:][]{Koyama2007} onboard {\it Suzaku}, we focus on front-illuminated CCDs (XIS0 and XIS3) which have slightly better spectral resolution than the back-illuminated CCD (XIS1).  We generated response files for the {\it XMM-Newton} Reflection Grating Spectrometer \citep[RGS:][]{denHerder2001}, taking account of energy-dependent spatial structures in the source, as we did in our previous analyses of the Galactic SNR Puppis~A \citep{Katsuda2012,Katsuda2013}.  The first-order RGS spectral resolution, which is determined by the size of the remnants in our case, is roughly $E/\Delta E \sim$25 and 180 for Kepler's SNR and SNR~0509-67.5, respectively.  This resolution is higher than that of the XIS of $E/\Delta E \sim$15.  

Backgrounds (BGs) were taken from source-free regions in the same fields of view.  The exception is the RGS spectrum for Kepler's SNR, for which the spatial extent of the source-free areas is limited; hence we used blank-sky data (Obs.~ID: 0051940401) taken immediately after the observation of Kepler's SNR.  Each spectrum was grouped into bins with at least 20 counts in order to allow us to perform a $\chi^2$ test.

\section{Ejecta Masses from Integrated X-Ray Spectra}

\subsection{Glancing at {\it Suzaku}'s XIS Spectra of the Three SNRs}

\begin{figure}
\begin{center}
\includegraphics[angle=0,scale=0.6]{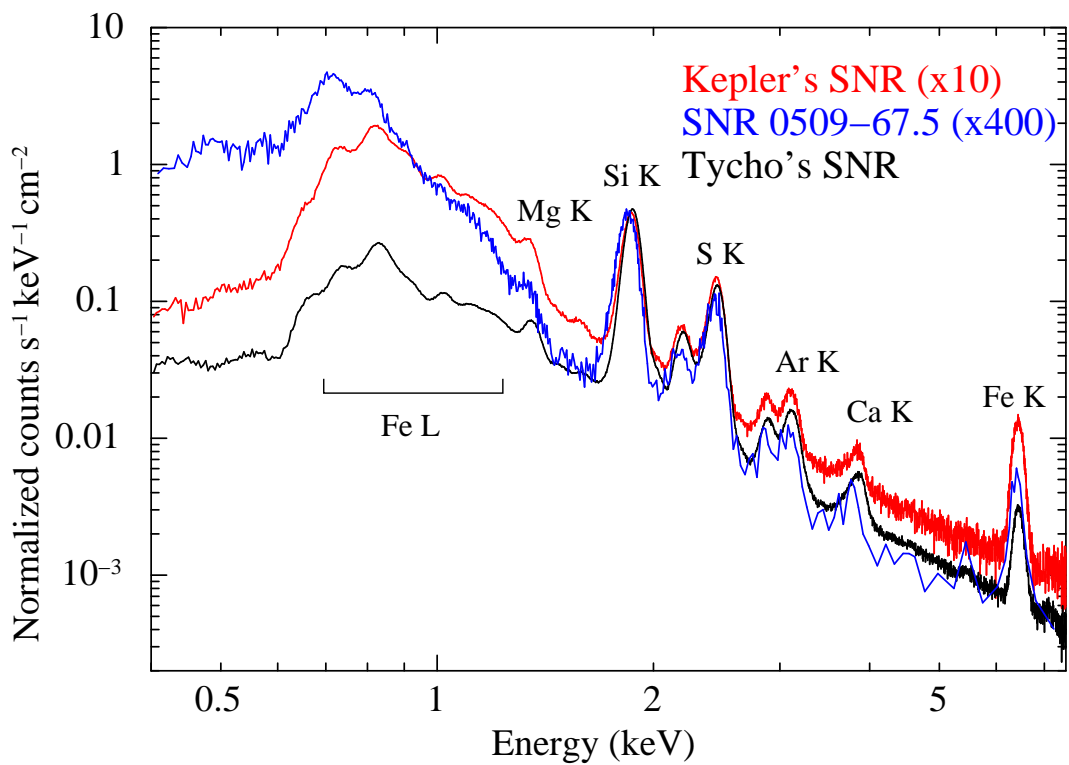}\hspace{1cm}
\caption{{\it Suzaku} XIS spectra of Kepler's SNR in red, Tycho's SNR in black, and SNR~0509-67.5 in blue.  The intensities are normalized such that the peak values of Si K-shell lines are equalized.  The error bars are omitted for clarity.  The Fe L and Fe K lines for Kepler's SNR and SNR~0509-67.5 are clearly stronger than those of Tycho's SNR.  It should be noted that these spectra are raw data, and are not corrected for the interstellar absorption.  
} 
\label{fig:spec_comp}
\end{center}
\end{figure}

We first examine the wide-band X-ray spectra for the three SNRs.  Figure~\ref{fig:spec_comp} shows integrated {\it Suzaku} XIS spectra of the SNRs, where the intensities are normalized to equalize the Si He$\alpha$ line.  A number of emission lines are apparent, mostly arising from the shocked SN ejecta.  Obviously, the strengths of Fe lines (both L- and K-shell transitions) are quite different among the three sources, whereas K-shell lines from intermediate mass elements (IME: Si, S, Ar, and Ca) are similar to one another, in the sense that shock-heated IGE-to-IME mass ratios for Kepler's SNR and SNR~0509-67.5 are significantly larger than that for Tycho's SNR.  It is reasonable to consider that the amount of unshocked cold ejecta in the three SNRs is similar among the remnants, based on their similar ages ($\sim$400\,yr), close evolutionary states \citep{Dickel1988,Reynoso1997,Katsuda2008,Vink2008,Katsuda2010}, and the similar reverse shock positions ($\sim$70\% of the forward shock radius) as we describe below.  Therefore, Kepler's SN must have produced a relatively large mass of IGE, similar to SNR~0509-67.5.  When combined with information from light-echo spectra, i.e., overluminous for SNR~0509-67.5 and normal for Tycho's SNR, we argue that Kepler's SN was likely an overluminous Ia event.  

This argument will be substantiated by our spectral fitting in the following section.  There are reasons why spectral fitting is required to estimate the ejecta masses quantitatively and correctly.  For example, while SNR~0509-67.5 has the strongest apparent Fe L lines below 1\,keV in the raw XIS spectra in Fig.~\ref{fig:spec_comp}, these lines are considerably weaker than those of Kepler's SNR, if we correct for the interstellar absorption.   Also notable is the slight shift of the line center energies of SNR~0509-67.5 with respect to those of the two Galactic SNRs.  This shift is due not to artificial/instrumental effects, but to a mixture of bulk motions and ionization effects, as we will show below.  Since the ionization state is sensitive to X-ray line intensities, we need to care about it in estimating ejecta masses.  All of these effects can be accounted for through spectral fitting. 

\subsection{Quantitative Mass Estimates Based on Spectral Fitting}

\begin{figure}
\begin{center}
\includegraphics[angle=0,scale=0.45]{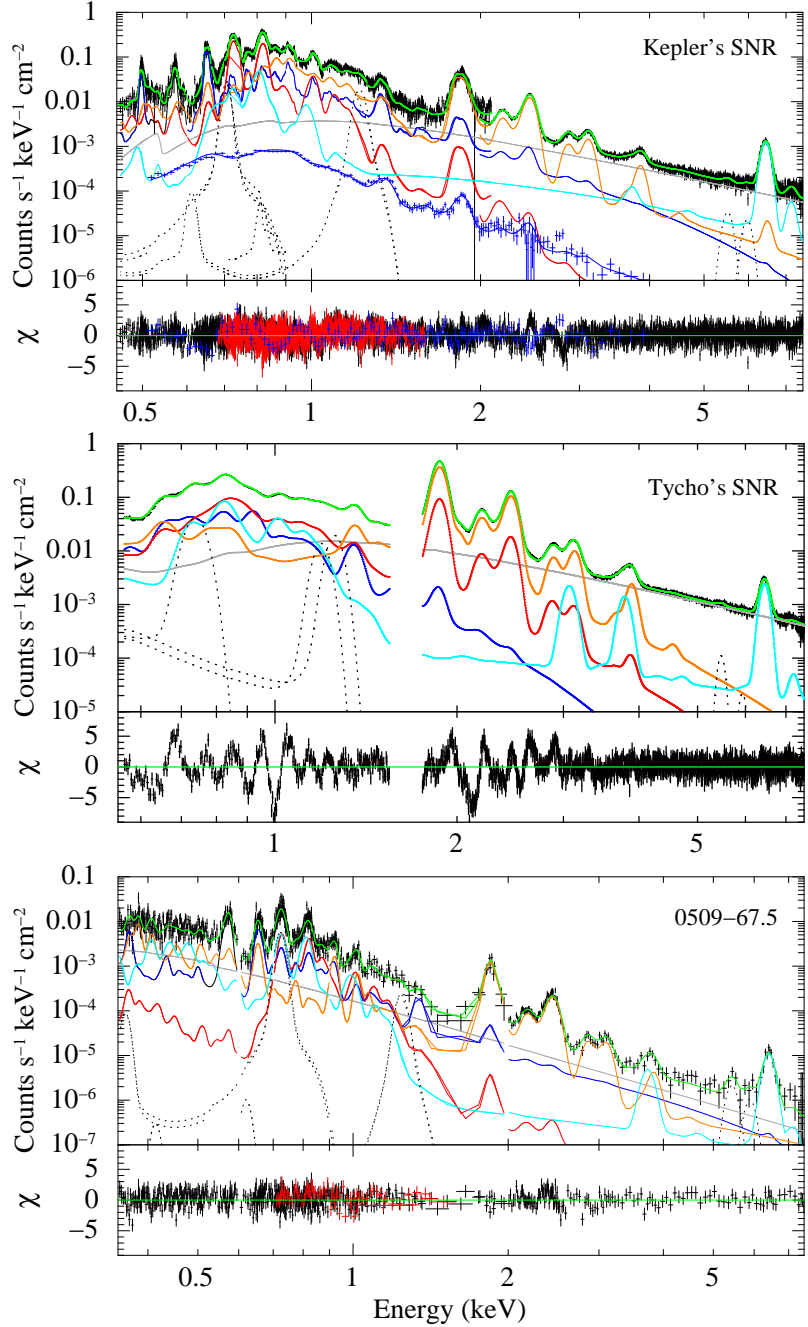}\hspace{1cm}
\caption{Spatially integrated X-ray spectra for Kepler's SNR (upper panel), Tycho's SNR (middle panel), and SNR~0509-67.5 (lower panel).  We use the {\it XMM-Newton} RGS below 2\,keV and the {\it Suzaku} XIS above 2\,keV for both Kepler's SNR and SNR~0509-67.5, but only the {\it Suzaku} XIS for Tycho's SNR.  For Kepler's SNR, we simultaneously fit the {\it Chandra} ACIS spectrum extracted from the CSM knots (Knots 1 and 2 in Fig.~\ref{fig:kepler_image}).  The lower panels show residuals.  Individual components, i.e., swept-up, three ejecta, and power-law, are separately shown in blue, red+orange+light blue, and gray, respectively.  
} 
\label{fig:spec_whole}
\end{center}
\end{figure}

In our spectral fitting, we use {\it XMM-Newton}'s RGS below 2\,keV and {\it Suzaku}'s XIS above 2\,keV for Kepler's SNR and SNR~0509-67.5.  On the other hand, for Tycho's SNR, we use only data from {\it Suzaku}'s XIS, since the ``slitless" RGS cannot provide a high-resolution spectrum due to the relatively large angular size of the remnant (diameter of $\sim$8.5$^{\prime}$).  Spatially-integrated X-ray spectra for the three SNRs are shown in Fig.~\ref{fig:spec_whole}.  In these plots, only the first-order RGS spectra are shown, but we use both first and second orders when modeling the data.  The data points in blue for Kepler's SNR is a local-BG subtracted CSM spectrum extracted from the summed-up regions of knots 1 and 2 (see, Section~4), by using {\it Chandra}'s Advanced CCD Imaging Spectrometer \citep[ACIS:][]{Garmire2003}.  We fit this spectrum with a CSM-only model having all normalizations of ejecta components fixed to zero, so that we can separate out the CSM emission as cleanly as possible.  Note that, for the spectrum of Tycho's SNR, we exclude data below 0.6\,keV owing to poor signal-to-noise ratios due to a relatively heavy interstellar absorption together with the low energy tail emission from higher energy lines \citep{Koyama2007}, and in the 1.55\,keV--1.75\,keV band due to a Si-edge calibration concern with the XIS\footnote{http://heasarc.gsfc.nasa.gov/docs/suzaku/analysis/sical.html}.

We fit these spatially-integrated spectra with an absorbed, {\tt vpshock} + three {\tt vnei}s + {\tt power-law} + several Gaussian components model in the XSPEC package \citep{Arnaud1996}.  Here, the {\tt vpshock} and the {\tt power-law} components are supposed to be the thermal and synchrotron radiation, respectively, from the swept-up medium.  The three {\tt vnei} components represent emission from SN ejecta.  At least three {\tt vnei} components are required from a statistical point of view (F-test probability of $>$99\%); we chose to not add more in order to keep our model as simple as possible.  The several Gaussian components represent lines of Fe L and/or O K, Fe L and/or Ne K, Cr K, and Mn K not well represented in the broad band models.  

While details of our parameter treatment can be found in Table~\ref{tab:param}, we elaborate some of them below.  We fix the abundance of either Si/H or Fe/H to be 10$^{5}$ times the solar value, assuming that the three {\tt vnei} components are nearly pure metal plasmas.  Most of the emission from IME is reproduced by two {\tt vnei} components, for which the two-components nature may reflect a temperature/density gradient within the shocked ejecta; e.g., a hydrodynamical model of Type Ia SNR evolution in a uniform ambient density predicts a strong temperature decrease (as well as density increase) from the reverse shock to the contact discontinuity \citep{Dwarkadas1998}.  For these components, we tie the metal abundances to each other except for Fe (and Ni).  The remaining {\tt vnei} component is introduced to reproduce most of the Fe K emission.  We assume the Ni/Fe abundance ratio to be that expected in the classical deflagration model \citep[W7:][]{Nomoto1984}; the assumed value of Ni/Fe abundance does not affect our spectral fitting, since Ni lines are not clearly detected in the spectra.  In this Fe-rich component, we allow abundances of Ar and Ca to vary freely; otherwise we see significant residuals at low-energy sides of the K lines of these elements, especially for Tycho's SNR.  This would indicate that some Ar and Ca may be mixed into IGE-rich layers, although we cannot rule out other possibilities, e.g., the presence of strongly redshifted components for these elements.  In any case, the ejecta masses will not change dramatically, hence we here adopt the simplest solution in terms of the spectral modeling.  

We fix the absorbing column densities at $N_\mathrm{H} = 6.4\times10^{21}$\,cm$^{-2}$, 1$\times10^{22}$\,cm$^{-2}$, and $6\times10^{20}$\,cm$^{-2}$ adopting elemental abundances given by \citet{Wilms2000} for Kepler's SNR, Tycho's SNR, and SNR~0509-67.5, respectively.  The values for the two Galactic SNRs are determined by fitting synchrotron-dominated spectra from the outermost thin rims obtained by the {\it Chandra}'s ACIS, while that for SNR~0509-67.5 is taken from the literature \citep[e.g.,][]{Warren2004,Williams2011}.  Free parameters are the electron temperature, $kT_\mathrm{e}$, the ionization timescale, $n_\mathrm{e} t$, the normalization defined as $\int n_\mathrm{e} n_\mathrm{H} dV$ where $n_\mathrm{e}$ and $n_\mathrm{H}$ are electron and hydrogen densities and $V$ is the emitting volume, and the redshift of each component, some of which are linked among different components in order to better constrain the parameters.  Photon indices are allowed to vary freely.  In addition, we multiply the {\tt gsmooth} model to each thermal component in order to take account of line-broadening effects, for which we assume the widths to be proportional to the line energies.  

As shown in Fig.~\ref{fig:spec_whole}, the best-fit models in green represent the data well. Individual components are shown in blue (swept-up medium), red (ejecta1 = the lower-temperature IME-dominated component), orange (ejecta2 = the higher temperature IME-dominated component), light blue (ejecta3 = the Fe-rich component), gray (power-law), and dotted black (Gaussians).  Note for Kepler's SNR that the swept-up components seen for both {\it XMM-Newton}'s RGS and {\it Chandra}'s ACIS have the same intrinsic spectral shapes, but look different due to distinct spectral responses.  The lower panels in Fig.~\ref{fig:spec_whole} show residuals; the red points represent the RGS second order data that are not shown in the upper panels.  The reduced-$\chi^2$ values summarized in Table~\ref{tab:param} are 1.38 (SNR~0509-67.5), 1.45 (Kepler's SNR), and 2.91 (Tycho's SNR).  The relatively large value for Tycho's SNR is not surprising, given that it is the brightest source of the three SNRs and the larger spectral bin widths (3.65\,eV at the minimum using the XIS) than those of the other two sources ($<$1\,eV using RGS), making systematic uncertainties relatively serious for this SNR.  We do not, however, introduce systematic uncertainties in our spectral fitting, since they do not affect our main conclusions.

We check the robustness of our spectral analysis by comparing the best-fit parameters in Table~\ref{tab:param} with previous measurements.  First of all, the N abundance of the swept-up ({\tt vpshock}) component in Kepler's SNR is quantitatively consistent with previous optical measurements \citep{Blair1991}, showing that this component is the CNO-processed stellar wind material, i.e., the CSM, as we expected.  Also, the lines in the CSM component (mainly N Ly$\alpha$, O He$\alpha$, and O Ly$\alpha$) are significantly narrower than those of ejecta components, assuring robustness of the spectral separation between the CSM and the ejecta.  Unfortunately, the data do not allow us to measure N abundances or line widths of the CSM components for the other two SNRs because of the relatively poor spectral resolution (Tycho's SNR) and statistics (SNR~0509-67.5).  As for the photon indices, the best-fit values agree with previous X-ray measurements \citep[e.g.,][]{Tamagawa2009,Cassam-chenai2007,Warren2004,Kosenko2008}.  In addition, by converting the line broadening of the ejecta component to bulk velocities, we obtain $v_\sigma$ = 2740$^{+40}_{-60}$\,km\,s$^{-1}$, 2970$^{+20}_{-10}$\,km\,s$^{-1}$, and 3570$^{+200}_{-230}$\,km\,s$^{-1}$ for Kepler's SNR, Tycho's SNR, and SNR~0509-67.5, respectively, where the velocities include effects of both thermal broadening and shell expansion whose contributions are hard to distinguish from our data.  These are somewhat smaller than previous estimates for Tycho's SNR \citep{Hayato2010} and SNR~0509-67.5 \citep{Kosenko2008}, which might be explained by multiple ionization effects in our analysis.

\begin{figure}
\begin{center}
\includegraphics[angle=0,scale=0.6]{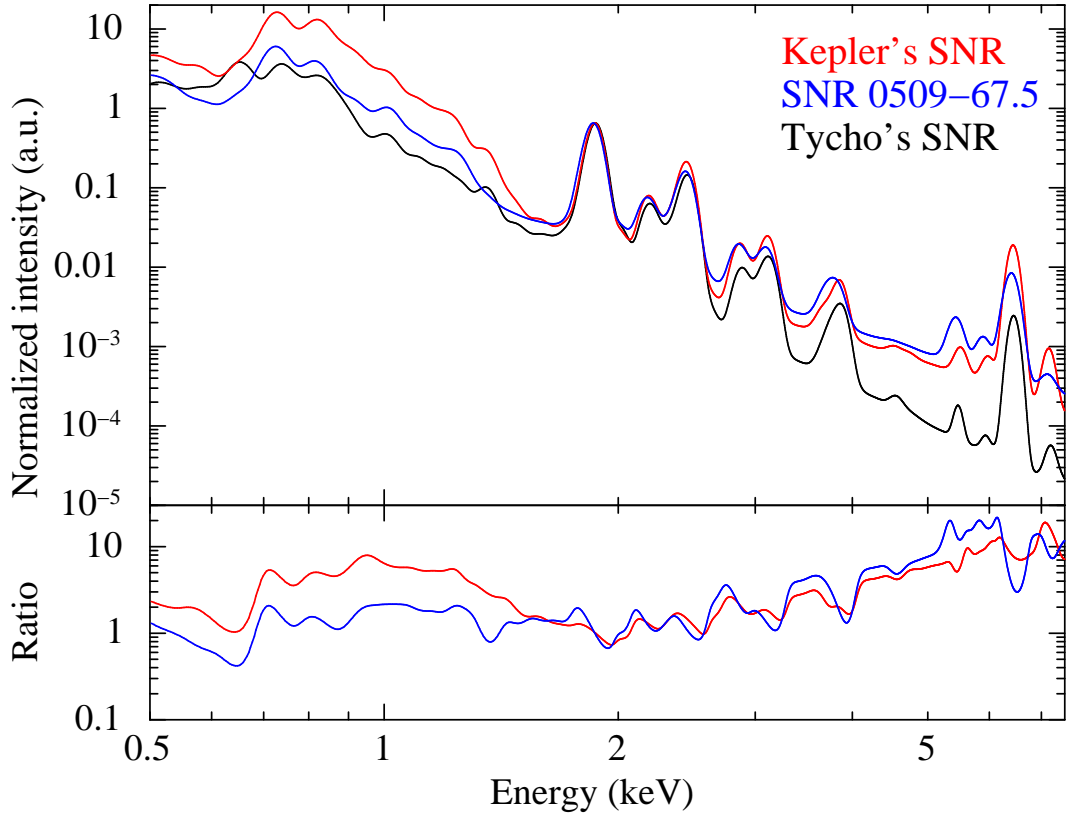}\hspace{1cm}
\caption{Comparison of unabsorbed ejecta emission models for Kepler's SNR in red, SNR~0509-67.5 in blue, and Tycho's SNR in black.  The models are convolved with the XIS response function, and the intensities are normalized such that the peaks of Si K-shell lines are equalized.  The lower panel shows spectral ratios with respect to Tycho's SNR.  
}
\label{fig:ejecta_comp}
\end{center}
\end{figure}

We next look into details of ejecta components.  Figure~\ref{fig:ejecta_comp} illustrates unabsorbed spectra of the total ejecta components with intensities normalized at Si K-shell lines, in order to clearly reveal the differences among the three SNRs.  We can readily see that the Fe K and L lines for Kepler's SNR and SNR~0509-67.5 are much stronger than those for Tycho's SNR.  This is also evident from the spectral ratios in the lower panel.  We then quantitatively estimate ejecta masses, using the best-fit parameters listed in Table~\ref{tab:param}.  As the first step, we calculate $n_\mathrm{e}$/$n_\mathrm{H}$ for each component, taking account of the elemental abundances and the ionization balance \citep[in the SPEX code:][]{Kaastra1996} at the $kT_\mathrm{e}$ and $n_\mathrm{e}t$ values measured.  With this information, we can calculate $\int n_\mathrm{H}^2 dV$ which can be converted to $\int n_\mathrm{X}^2 dV$ by multiplying each elemental abundance.

\begin{figure}
\begin{center}
\includegraphics[angle=0,scale=0.6]{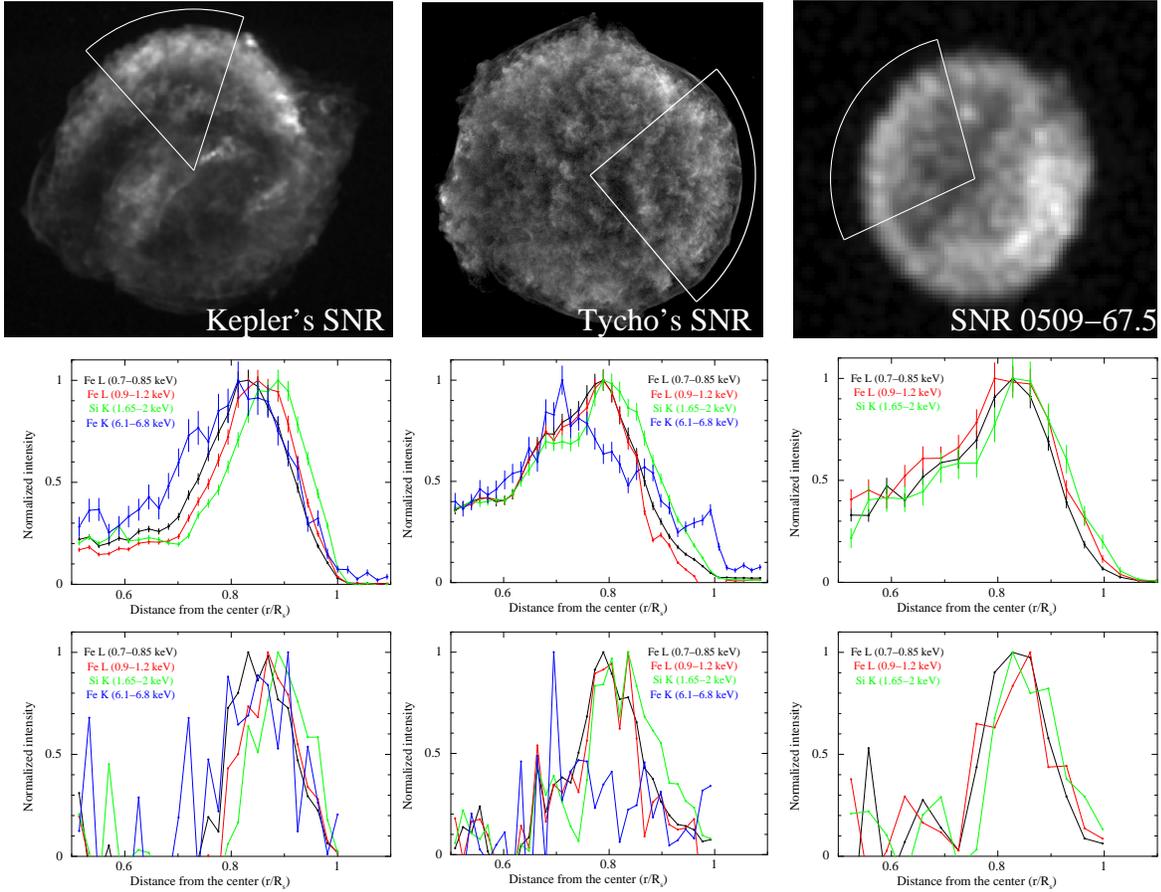}\hspace{1cm}
\caption{Top panels: {\it Chandra} ACIS images of the three SNRs analyzed in this paper.  We extract radial profiles from the white pie-shaped areas covering relatively smooth, limb-brightened regions.  
Middle panels: X-ray radial profiles from the pie-shaped regions in top panels.  The profiles are generated in four different energy bands: 0.7--0.85\,keV (black), 0.9--1.2\,keV (red), 1.65--2.0\,keV (green), and 6.1--6.8\,keV (blue).  
Bottom panels: Deprojected radial profiles corresponding to the profiles in the middle panel.
} 
\label{fig:rad_dep_prof}
\end{center}
\end{figure}

To deduce densities (and masses) from $\int n_\mathrm{X}^2 dV$, we need to know emitting volumes of individual components.  To this end, we generate deprojected radial profiles from high-angular resolution data obtained with the {\it Chandra}'s ACIS.  We first extract energy-band images for 0.7--0.85\,keV (Fe L), 0.9--1.2\,keV (Fe L), 1.65--2.0\,keV (Si K), 2.25--2.6\,keV (S K), 2.95--3.25\,keV (Ar K), 3.65--4.1\,keV (Ca K), 4.0--6.0\,keV (continuum), and 6.1--6.8\,keV (Fe K).  For each band image, we subtract continuum emission extrapolated according to the power-law slopes given in Table~\ref{tab:param}.  We then generate continuum-subtracted radial profiles, by focusing on a smooth limb-brightened region of each SNR, i.e., north, west, and northeast for Kepler's SNR, Tycho's SNR, and SNR~0509-67.5, respectively, as shown in Fig.~\ref{fig:rad_dep_prof} upper panels.  The radial profiles produced in this way for four selected energy bands are displayed in the middle panels of Fig.~\ref{fig:rad_dep_prof}.  We see a general trend, with Fe K, Fe L, and Si K peaking at increasing radius, consistent with a previous {\it XMM-Newton} study for Kepler's SNR \citep{Cassam-chenai2004}.  In addition, we find that the peak position of Ne-like Fe L lines (0.7--0.85\,keV) is located systematically inside that of the more ionized Fe L lines (0.9--1.2\,keV).  This fact suggests that in each remnant, the Fe is indeed primarily SN ejecta heated by a reverse shock propagating toward the center.  These radial profiles are deprojected by using the ``onion-peeling" method, which estimates the emission at each radius repeated removal of the contribution from concentric outer shells \citep[e.g.,][]{Fabian1980}.  The deprojected profiles are displayed in the lower panels in Fig.~\ref{fig:rad_dep_prof}, where we do not give error bars as they strongly depend on the geometry of the ejecta, but the values should be $\lesssim\pm$10\% if the assumed spherical symmetric distribution of the ejecta is correct.

These deprojected profiles are used to estimate emitting volumes of the individual {\tt vnei} components.  Since the profiles of Kepler's SNR look similar to those of SNR~0509-67.5, we assume that the emitting geometries are the same.  The reverse shock radii, $R_\mathrm{RS}$, are taken to be the sharp rises of the Fe K (and Ne-like Fe L), which are 0.7 and 0.65 times the forward shock radii, $R_\mathrm{FS}$, for Kepler's SNR and Tycho's SNR, respectively, where we define $R_\mathrm{FS}$ from eye inspections to be 107$^{\prime\prime}$, 256$^{\prime\prime}$, and 14.8$^{\prime\prime}$ for Kepler's SNR, Tycho's SNR, and SNR~0509-67.5, respectively.  We note that the value of $R_\mathrm{RS}$ found in this way for Tycho's SNR is between those from a previous {\it Chandra} study \citep{Warren2005} and a recent measurement from Fe K$\beta$'s radial profiles with {\it Suzaku} \citep{Yamaguchi2014}.  The hot, low-ionization, Fe-rich plasma (ejecta 3 in Table~\ref{tab:param}) resides in a layer from $R_\mathrm{RS}$ to 0.85 (0.75) $R_\mathrm{FS}$, i.e., the inner radius of the half maximum of Si K's deprojected profiles for Kepler's SNR (Tycho's SNR).  Other IME-rich components, i.e., ejecta2 and ejecta3 in Table~\ref{tab:param}), occupy a shell outside the inner Fe-rich layers.  The outer extent of the IME-rich components for Kepler's SNR (Tycho's SNR) is taken to be 0.97 (0.93) $R_\mathrm{FS}$, which corresponds to the position where we see a sharp rise of the deprojected Si K profile for Kepler's SNR (Tycho's SNR).  We note that the value for Tycho's SNR represents the contact discontinuity suggested from a previous {\it Chandra} study \citep{Warren2005}.  We allocate volume filling factors for the two IME-rich {\tt vnei} components, such that their electron pressures are equalized: ($f_\mathrm{ej1}$ : $f_\mathrm{ej2}$) = (0.05 : 0.95), (0.2 : 0.8), and (0.015 : 0.985) for Kepler's SNR, Tycho's SNR, and SNR~0509-67.5, respectively.  We then compute densities and masses for individual elements in each ejecta component.  

We note that the surface brightness of Kepler's SNR is quite asymmetric; the northern rim is much brighter than the southern rim.  This could lead to biased mass estimates, if we use only one side (either north or south) of the rim in estimating emitting volumes.  Therefore, we examine radial profiles of the southern portion in addition to those of the northern portion, finding that the reverse shock position as well as the IGE and IME layers are close to those of Tycho's SNR rather than the northern portion of Kepler's SNR.  However, this modified emitting volumes result in only $\sim$10\% variations of ejecta masses toward increasing/decreasing direction for IGE/IME.  Such slight differences will not change our conclusion given below.

\begin{figure}
\begin{center}
\includegraphics[angle=0,scale=0.6]{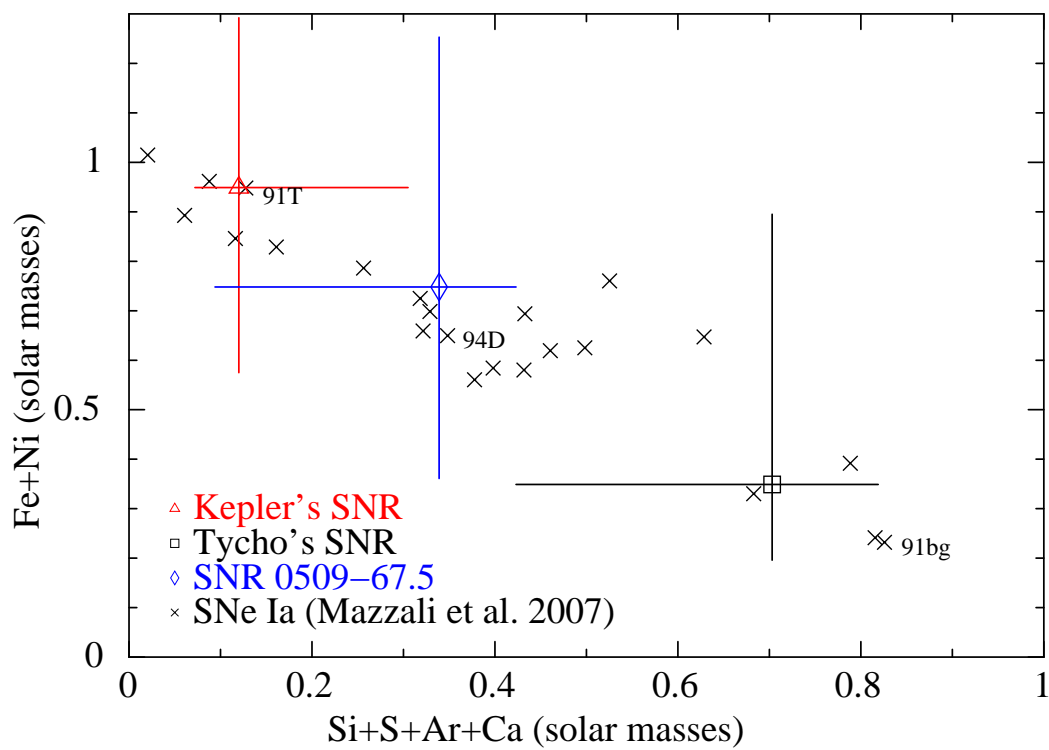}\hspace{1cm}
\caption{IGE masses vs.\ IME masses for 23 extragalactic SNe Ia \citep[crosses:][]{Mazzali2007}.  The data include a variety of subclasses of SNe Ia, i.e., overluminous (e.g., SN~1991T), normal (e.g., SN~1994D), and subluminous (e.g., SN~1991bg).  Our data for Kepler's SNR, Tycho's SNR, and SNR~0509-67.5 are shown as a red triangle, a black box, and a blue diamond, respectively.  The errors quoted are taken from Table~\ref{tab:ejecta_mass}.  
} 
\label{fig:ige_vs_ime}
\end{center}
\end{figure}

The estimates of integrated masses for the three ejecta components are summarized in Table~\ref{tab:ejecta_mass}.  We find that IGE masses for SNR~0509-67.5 and Kepler's SNR are close to each other, but are higher than that for Tycho's SNR.  This result quantitatively confirms our expectation (from the spectral comparison in Figs.~\ref{fig:spec_comp} and \ref{fig:ejecta_comp}) that Kepler's SN was an overluminous (91T-like) Ia event.  This is also consistent with the result drawn by hydrodynamical modeling of the X-ray spectrum \citep{Patnaude2012}.  A more convincing plot can be found in Fig.~\ref{fig:ige_vs_ime}, in which we plot our data points on the IGE--IME mass diagram of 23 extragalactic SNe Ia \citep{Mazzali2007} as in Fig.~\ref{fig:ige_vs_ime}, where several 91T-like SNe are clustered in the upper left corner, a few subluminous (or 91bg-like) objects are distributed in the lower right corner, and many normal Ia objects are located in between them.  We can see that Kepler's SNR is well within the overluminous (91T-like) category, while Tycho's SNR is among the normal SNe Ia.  In this plot, SNR~0509-67.5 can be found in a marginal region between overluminous and normal Ia events.  Threfore, this plot shows that the amount of IGE of Kepler's SNR is even larger than that of the confirmed overluminous Ia SNR~0509-67.5, strongly supporting the idea that Kepler's SN was an overluminous Ia event.

It should be noted in Table~\ref{tab:ejecta_mass} that (1) we provide fairly conservative uncertainties which are much larger than those from the statistical uncertainties in Table~\ref{tab:param}; (2) we list unshocked IGE masses, based on SN Ia nucleosynthesis models \citep{Nomoto1984,Maeda2010} combined with the reverse-shock positions estimated above; and (3) we assume that the total IME + IGE mass becomes $\sim$1.05\,M$_\odot$ \citep{Mazzali2007} by adjusting the distances to the remnants.  The details are described below.

To evaluate the errors, we consider a wider range of the electron temperature for the hot Fe-rich component than the statistical uncertainties, given possible systematic uncertainties on Fe L emissivities.  In fact, even if we fix $kT_\mathrm{e}=$2\,keV and $kT_\mathrm{e}=$15\,keV for this component, we derive similar quality fits; reduced-$\chi^2$ values for $kT_\mathrm{e}=$2\,keV and 15\,keV are, respectively, 1.54 and 1.49 for Kepler's SNR, 3.55 and 3.02 for Tycho's SNR, and 1.42 (for both 2\,keV and 15\,keV) for SNR~0509-67.5.  Thus, we consider these extreme temperature cases to be viable, and they provide the major source of the large uncertainties in Table~\ref{tab:ejecta_mass}.  

\begin{figure}
\begin{center}
\includegraphics[angle=0,scale=0.5]{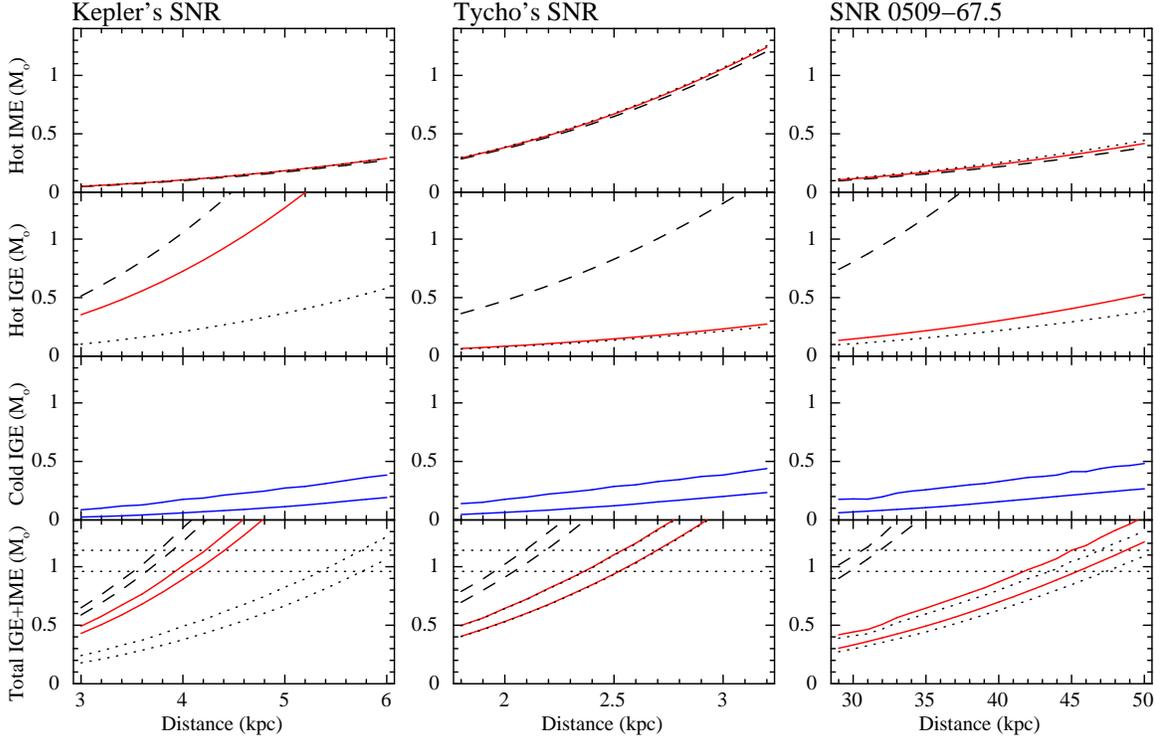}\hspace{1cm}
\caption{Ejecta masses as a function of distances to Kepler's SNR (left column), Tycho's SNR (central column), and SNR~0509-67.5 (right column).  The top-row, second-row, third-row, and the bottom-row displays masses for shocked IME (Si+S+Ar+Ca), shocked IGE (Fe+Ni), unshocked IGE, and their sum, respectively.  The dashed and dotted curves are obtained by fixing the electron temperature of the Fe-K emitting plasma at 2\,keV and 15\,keV, respectively, whereas solid red curves are responsible for the best-fit results.  The cold IGE masses are based on two nucleosynthetic models, i.e., W7 \citep[the upper curve:][]{Nomoto1984} and O-DDT \citep[the lower curve:][]{Maeda2010}.  These two cases result in the pairs of curves in the bottom panel.  In the bottom panels, the horizontal lines indicate the range of 0.96--1.14\,M$_\odot$ \cite{Mazzali2007}.
}
\label{fig:mass_vs_dist}
\end{center}
\end{figure}

The upper two rows of Figure~\ref{fig:mass_vs_dist} show the masses of shocked IME and IGE as a function of distance for three cases:  the best fits in the red solid curves; $kT_\mathrm{e} = 2$\,keV in the black dashed curves; and $kT_\mathrm{e} = 15$\,keV in the black dotted curves.  The third row shows unshocked ejecta masses for two SN Ia abundance distributions: the classical W7 model \citep[the upper curve:][]{Nomoto1984} and a two-dimensional (azimuthally-averaged) delayed-detonation model, i.e., off-center ignition O-DDT model \citep[the lower curve:][]{Maeda2010}.  While both of the two models are not responsible for overluminous Ia SNe but for normal Ia SNe, the amount and composition of the unshocked ejecta would not be different by much, since the abundance distributions of the two models examined here look (more) similar to those of overluminous cases rather than subluminous cases by \citet{Seitenzahl2013}.  To estimate the cold ejecta masses, we integrate the ejecta masses from $v_\mathrm{ej}=0$ up to $v_\mathrm{ej} = R_\mathrm{rs}$/age (the free expansion velocity at the reverse shock).  We find that the composition of the unshocked ejecta is almost pure IGE for both models, hence we name the y-axis label as ``cold IGE."  The bottom panels show the total masses (hot IME + hot IGE + cold IGE).  

In principle, the ejecta masses can take any values (even unrealistic ones), by changing the temperature and/or distances (see Fig.~\ref{fig:mass_vs_dist}).  Therefore, we introduce one constraint that the total masses of IGE plus IME are within 0.96--1.14\,M$_\odot$, according to a systematic analysis of 23 samples of well-observed extragalactic SNe Ia \citep{Mazzali2007}.  In the bottom panels, this mass range is given as horizontal dotted lines.  Limiting the total masses in this range, we can constrain distances to the remnants for each temperature case.  The distances for the best-fit cases (in between the two solid red lines) are $\sim$4.2\,kpc, $\sim$2.55\,kpc, and $\sim$46\,kpc for Kepler's SNR, Tycho's SNR, and SNR~0509-67.5, respectively.  While the values for the two Galactic SNRs agree with some of previous estimates \citep{Sankrit2005,Katsuda2008,Vink2008,Chevalier1980,Albinson1986}, all of the distances are slightly smaller than the canonical values \citep{Kerzendorf2014,Hayato2010,Alves2004,Macri2006}.  In particular, the distance to SNR~0509-67.5 in the LMC is relatively well determined.  Thus, if we set the distance to the current best estimate of 48.1\,kpc \citep{Macri2006}, we obtain the total IME+IGE mass to be $\sim$1.2\,M$_\odot$ (see, Fig.~\ref{fig:mass_vs_dist}).  This is just slightly larger than the upper end (1.14\,M$_\odot$) expected for SNe Ia \citep{Mazzali2007}, and such a deviation would be still acceptable.  On the other hand, there is also a possibility that SNR~0509-67.5 whose absorption within the LMC is negligible is indeed located at $\sim$2\,kpc front side with respect to the main body of the LMC.  Since accurate measurements of the total masses and distances, which require significant additional analysis efforts, are beyond the scope of this paper, we leave this issue for future work based on high-resolution X-ray spectra which will be available with the upcoming {\it ASTRO-H} \citep{Takahashi2014}.  Keeping the total IME+IGE masses within 0.96--1.14\,M$_\odot$, we calculate allowed masses of the IME and IGE for the three temperature cases ($kT_\mathrm{e}=$2\,keV, 15\,keV, and the best-fit).  The resultant nominal values along with the uncertainties are given in Table~\ref{tab:ejecta_mass}.

\section{The Circumstellar Medium in Kepler's SNR}

\begin{figure}
\begin{center}
\includegraphics[angle=0,scale=0.7]{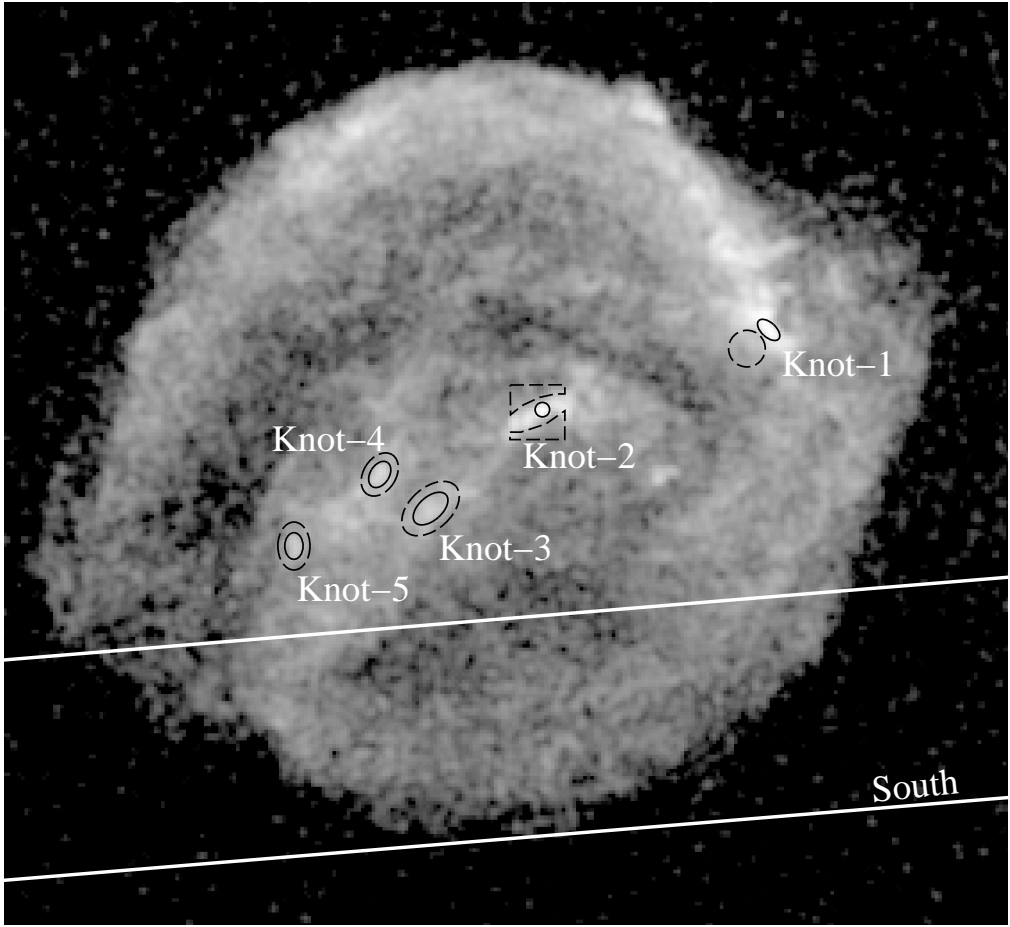}\hspace{1cm}
\caption{Spectral extraction regions for spatially-resolved CSM studies in Kepler's SNR.  For the diffuse CSM region, we extract the {\it XMM-Newton}'s RGS and {\it Chandra}'s ACIS spectra from the region within two white lines.  For the CSM knots, we extract {\it Chandra}'s ACIS spectra from a solid circle and four solid ellipses.  Dashed regions are where we extract local BGs.  
} 
\label{fig:kepler_image}
\end{center}
\end{figure}

\begin{figure}
\begin{center}
\includegraphics[angle=0,scale=0.6]{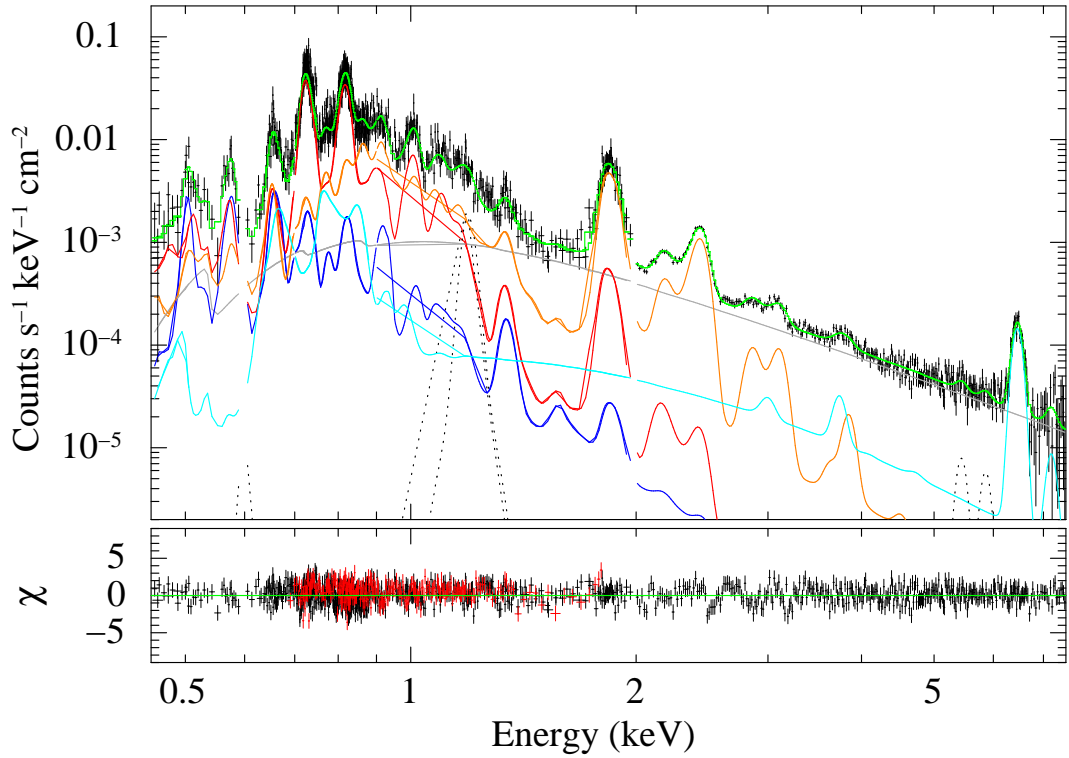}\hspace{1cm}
\caption{The combined RGS (below 2\,keV) and the {\it Chandra}'s ACIS (above 2\,keV) spectrum extracted from the southern portion of Kepler's SNR as shown in Fig.~\ref{fig:kepler_image}.  The spectrum is fitted with the same model as we applied to the integrated spectra in Fig.~\ref{fig:spec_whole}.  The lower panels show residuals.
} 
\label{fig:kepler_south_spec}
\end{center}
\end{figure}

We perform spatially-resolved spectral analyses to reveal details of the CSM, using the {\it XMM-Newton} RGS as well as the {\it Chandra} ACIS.  Our particular interest is to search for X-ray emission from the ``diffuse" CSM found by \citet{Williams2012} using {\it Spitzer} IR observations in the form of a lower temperature dust component, distinct from the hotter dust component associated with dense/knotty CSM.  To this end, we extract an RGS spectrum from the region within the solid lines in Fig.~\ref{fig:kepler_image}, where no CSM knots are present.  Figure~\ref{fig:kepler_south_spec} displays the RGS spectrum together with the best-fit model --- the same model as we applied to the integrated spectra in Fig.~\ref{fig:spec_whole}.  We find a line feature at $\sim$0.5\,keV, which can be modeled by N Ly$\alpha$ arising from the swept-up component.  The best-fit parameters are listed in Table~\ref{tab:param}; they show the presence of an enhanced N abundance in the swept-up component, similar to that found for the entire remnant (Figure~\ref{fig:spec_whole} and Table~\ref{tab:param}).  This is evidence that this component originates from the CNO-processed CSM, revealing that the diffuse CSM can be seen in X-rays as well as in the IR \citep{Williams2012}.

The diffuse CSM can be reasonably interpreted as material in steady outflow from the progenitor system, which refilled a cavity created by the major eruption a few 10$^{4}$\,yrs ago suggested by optical proper motions of the CSM knots \citep{Bandiera1991}.  The density and the mass of the diffuse CSM are calculated to be 2.8\,cm$^{-3}$ and 0.08\,M$_\odot$, respectively, at a distance of 4.2\,kpc, based on the assumption that the emitting region is a polar-cap shell of 0.97--1 $R_\mathrm{FS}$ with a filling factor of unity.  This density is consistent with that inferred from IR observations \citep{Williams2012}.  The total mass integrated over the entire remnant is $\sim$0.3\,M$_\odot$, given that the region we examined covers only $\sim$30\% of the entire remnant.  It should be noted that the total mass might increase by a factor of a few if we take into account the north-south asymmetry seen in both the IR and X-rays.  Assuming a wind speed of $v_\mathrm{w}=$10\,km\,s$^{-1}$, we estimate a mass-loss rate of 1.5$\times10^{-5}$($v_\mathrm{w}$/10\,km\,s$^{-1}$)\,M$_\odot$\,yr$^{-1}$, is needed to provide the estimated mass of 0.3\,M$_\odot$ within a radius of 2.1\,pc at a distance of 4.2\,kpc.  Such a large mass-loss rate suggests that the progenitor is an AGB star, which is consistent with results from recent hydrodynamical simulations \citep{Velazquez2006,Chiotellis2012,Toledo-roy2014}.  

\begin{figure}
\begin{center}
\includegraphics[angle=0,scale=0.6]{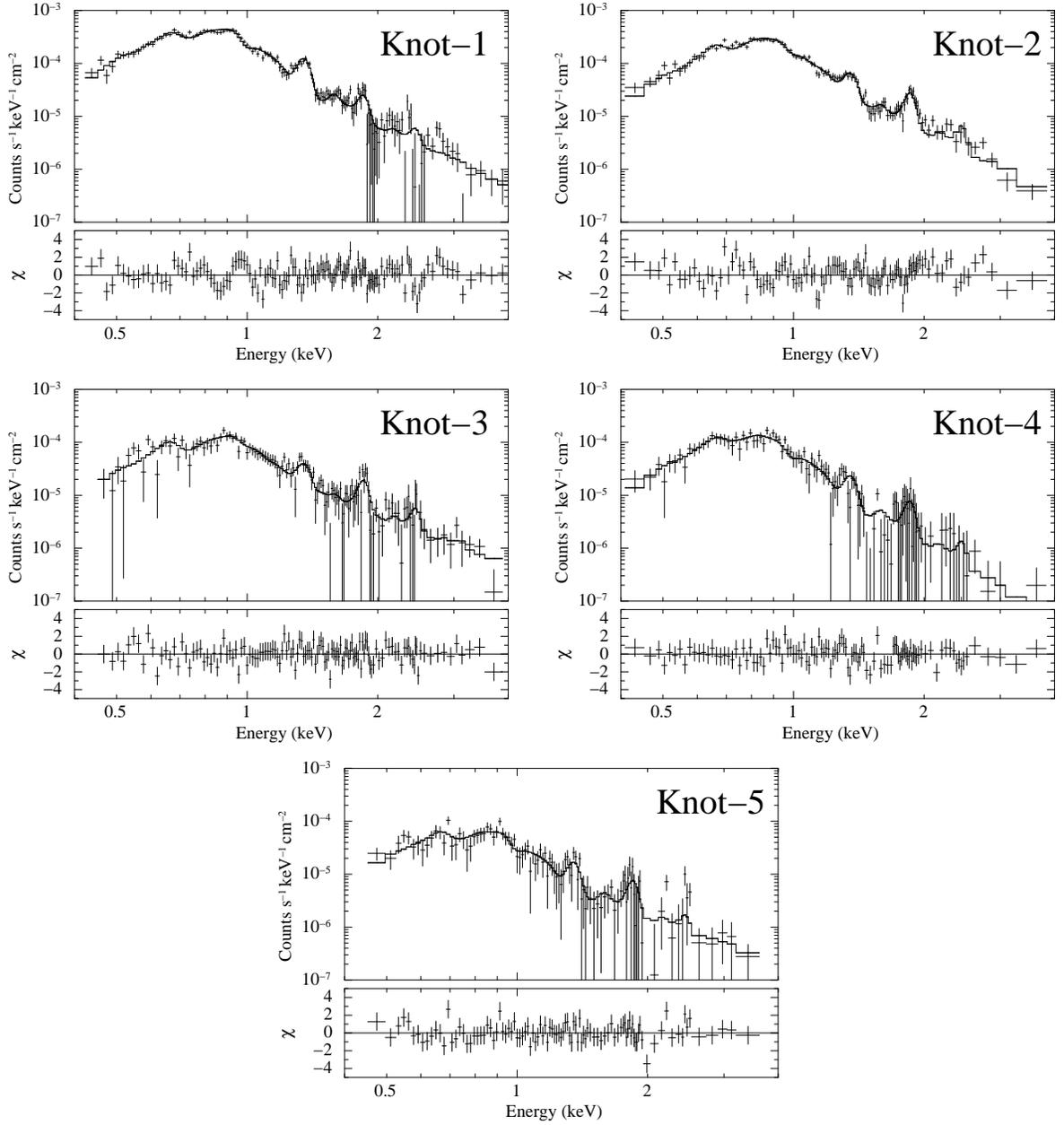}\hspace{1cm}
\caption{Local-BG subtracted {\it Chandra}'s ACIS spectra for CSM knots as shown in Fig.~\ref{fig:kepler_image}.  The spectra are fitted with a single-component {\tt vpshock} model.  The lower panels show residuals.
} 
\label{fig:kepler_chandra_spec}
\end{center}
\end{figure}

Next, we turn to the bright knotty features which are positionally coincident with optically bright knots.  We examine five such features (Knots 1--5) as shown in Fig.~\ref{fig:kepler_image}.  By subtracting local BGs (dashed regions next to the source regions), we extract clean CSM spectra.  The spectra are fitted by a single-component {\tt vpshock} model.  In the fitting procedure, the $kT_\mathrm{e}$, $n_\mathrm{e}t$, and normalization are treated as free parameters.  Abundances of O, Ne, and Mg are allowed to vary freely for the brightest Knots 1 and 2, while we fix all metal abundances to the solar values for the other relatively-dim knots.  The spectra with the best-fit models are shown in Fig.~\ref{fig:kepler_chandra_spec}.  The best-fit parameters are summarized in Table~\ref{tab:csm}.  The temperatures and ionization timescales are roughly consistent with those derived from the integrated RGS spectrum (Table~\ref{tab:param}) and previous {\it Chandra} measurements \citep{Reynolds2007}.  

To derive densities and masses, we assume the X-ray--emitting plasma's geometry to be spherical for Knot-2 and a rugby-ball shape for the others.  We also assume filling factors to be unity for all the regions.  The estimated densities and masses are listed in Table~\ref{tab:csm}.  The densities are an order of magnitude lower than the optical measurements \citep{Dennefeld1982}, suggesting that X-rays arise from relatively tenuous regions surrounding (or close to) optical dense cores.  The total mass of the five regions is 1.1$\times$10$^{-3}$\,M$_\odot$.  Clearly, these regions contain quite limited amounts of the total CSM knots, since the optical CSM knots are distributed more widely than the selected regions \citep[e.g.,][]{Bandiera1991}.  Indeed, the normalization summed-up for the five regions (see, Table~\ref{tab:csm}) is only $\sim$1/40 of the RGS-measured normalization integrated for the entire remnant (see, Table~\ref{tab:param}).  Part of the integrated normalization should be attributed to the diffuse CSM, whose fraction is estimated to be $\sim$20\%, based on the normalizations of the entire remnant and the southern part in Table~\ref{tab:param}.  Thus, we infer the total mass of the knotty CSM in the entire remnant to be $\sim$0.035\,M$_\odot$ (=1.1$\times$10$^{-3}$\,M$_\odot \times 40 \times 0.8$).  This would be slightly increased by adding optically emitting knots.  Such a large amount of the CSM is consistent with those observed in SNe Ia-CSM.  

Unlike Ia-CSM SNe, the dense/knotty CSM in Kepler's SNR is located far away from the progenitor star, as evidenced by three independent pieces of information.  First, the slow proper motions of the CSM knots measured in optical \citep{Bandiera1991} allow movements of only 5.3$^{\prime\prime}$ (or 0.1\,pc at a distance of 4.2\,kpc) after the explosion.  This is evidence that the knots were already $\sim$2\,pc (i.e., the SNR radius) away from the progenitor star at the time of the SN explosion.  Second, dividing the ionization timescale by the electron density in Table~\ref{tab:csm}, we infer the shock-heating time to be $\sim$50--200\,yrs.  This means that interactions between the CSM and the blastwave started at a very late phase, i.e., $\gtrsim$200\,yrs after the explosion.  Third, the light curve of Kepler's SN does not show any hints of early-phase CSM interactions.  Figure~\ref{fig:kepler_lc} shows two light curves of Kepler's SN obtained by European observers in black and Korean observers in red \citep[][]{Baade1943,Clark1977,Schaefer1996}.  For comparison, we show typical light curves of several kinds of SNe, i.e., Ia-CSM \citep[SN~2002ic:][]{Wood-vasey2004}, 91T-like \citep[SN~1991T:][]{Lira1998}, 91bg-like \citep[SN~1991bg:][]{Filippenko1992}, and Type IIp \citep[SN~1999em:][]{Elmhamdi2003}.  All of the light curves are normalized such that their peak magnitude is zero, in order to highlight the difference of their shapes.  We find that the light curves of Kepler's SN (both European and Korean versions) are clearly distinct from that of SN~2002ic (Ia-CSM), showing that there was no massive CSM in close proximity to the progenitor ($\lesssim$0.01\,pc).  This assertion would be fairly robust, given that both of the two independent historical light curves, generally agreeing with each other within 1 mag, decline quite fast ($\Delta$mag$\sim$5 at day$\sim$100 from the maximum light), which is in stark contrast to the flat light curves seen for SNe Ia-CSM.

\section{Discussion}

\begin{figure}
\begin{center}
\includegraphics[angle=0,scale=0.6]{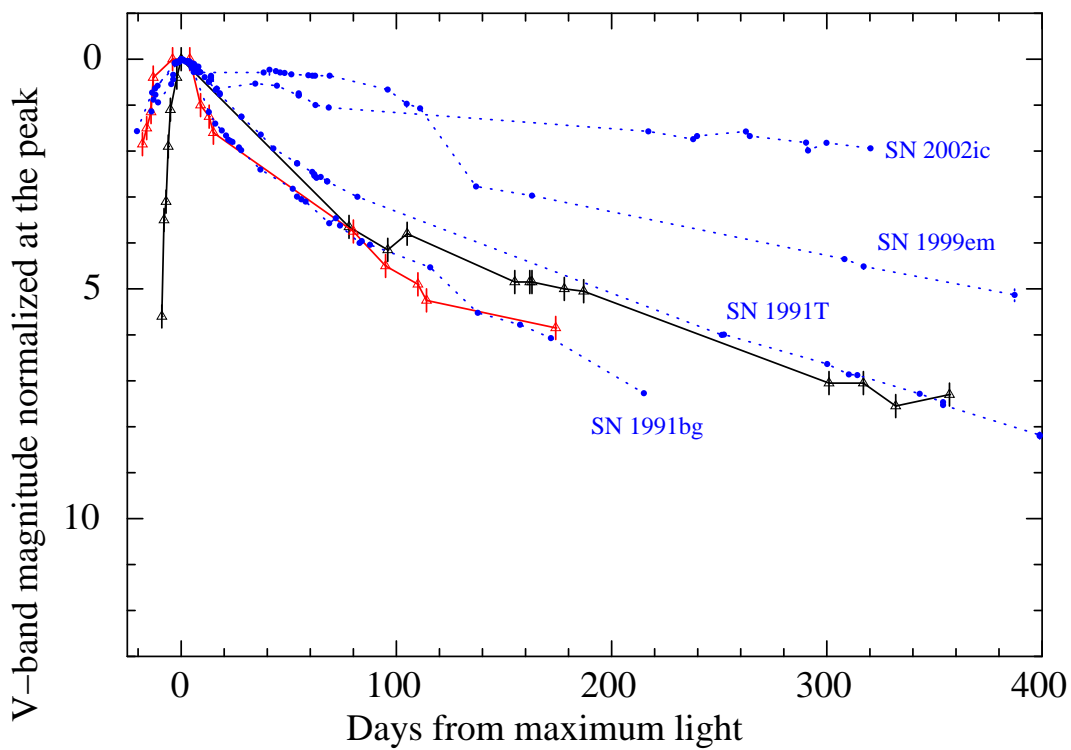}\hspace{1cm}
\caption{Light curves of Kepler's SN and other modern SNe.  The triangular data points in black and red tied by solid lines are responsible for Kepler's SN, obtained by European and Korean observers, respectively.  The errors are assumed to be $\pm$0.25 mag \citep{Baade1943}.  Other filled-circle data tied by dotted lines are responsible for recent SNe including Ia-CSM (SN~2002ic), 91T-like (SN~1991T), 91bg-like (SN~1991bg), and Type IIp (SN~1999em).   
} 
\label{fig:kepler_lc}
\end{center}
\end{figure}

Our work strongly suggests that Kepler's SN was likely to be a 91T-like event, based on the ejecta abundances --- the dominance of IGE.  Since detailed light curves of Kepler's SN were taken by both European and Korean observers in A.D.1604--1605, it is interesting to ask if these light curves are consistent with those of overluminous SNe Ia obtained by modern telescopes.  Unfortunately, we found it difficult to discriminate the overluminous and subluminous SNe Ia from the light-curve shapes.  As shown in Fig.~\ref{fig:kepler_lc}, the light curve measured by Johannes Kepler and some other European is best represented by that of SN~1991T, whereas the Korean version (red data in Fig.~\ref{fig:kepler_lc}) is closer to that of SN~1991bg (subluminous).  This discrepancy probably means that brightness uncertainties from naked-eye measurements are too large to discriminate among SNe Ia subtypes that follow the light-curve decline rate versus peak luminosity correlation \citep[i.e., the ``Phillips relation":][]{Phillips1999}.  Another important piece of information from the light curve is the peak brightness.  Assuming the absolute maximum brightness of Kepler's SN to be -19.5 mag, a typical value for 91T-like events, and a visual extinction to be 3.27 mag \citep{Schaefer1996}, we estimate the distance to Kepler's SNR to be either 6.3\,kpc or 4.5\,kpc from the apparent maximum brightness (-2.25 or -2.95) reported in Europe and Korea, respectively.  Therefore, the apparent maximum brightness from the Korean observations matches our result better.  We should, however, note that the distance of 6.3\,kpc from European maximum brightness is more consistent with recently preferred value of $\sim$6\,kpc.  Given these considerations, the only solid conclusion from the light curve studies is that Kepler's SN was different from Ia-CSM and Type IIp, but was close to Phillips-relation SNe Ia.


In addition, we confirmed the presence of dense X-ray emitting CSM in Kepler's SNR, and estimated its total mass in the entire remnant to be $\sim$0.035\,M$_\odot$.  This large mass reminds us of the massive CSM associated with SNe Ia-CSM.  However, as we discussed above, the light curve is inconsistent with a Ia-CSM event for Kepler's SN, suggesting the different distances between the CSM and the progenitor star; a few pc away from the progenitor for Kepler's SN is much farther than those seen in Ia-CSM events ($\lesssim$0.01\,pc).  Thus, we argue that Kepler's SN was likely to be a 91T-like event, and started to interact with the dense CSM $\gtrsim$200\,yrs after the SN explosion.  

In this way, Kepler's SNR has provided us with observational evidence for a recently proposed connection between 91T-like events and Ia-CSM objects: i.e., the two Ia subclasses are actually the same phenomenon with seeming differences due to different timing of the interaction with a massive CSM \citep{Leloudas2013}.  This speculation was originally inspired by optical observations of SN~2002ic and PTF11kx, whose optical spectra initially resembled a 91T-like template, but started to develop an H$\alpha$ line after 22 days and 59 days after explosions, respectively, evolving to more like a Type IIn template.  Since the timing of the CSM interaction is not unique, we can easily guess a possibility that the timing was too late to show signatures of the CSM, as the SNe fade after the explosions.  While such SNe would be classified as 91T-like events, if we could monitor them for several hundred years, we should be able to detect a signature of a massive CSM.  We believe that Kepler's SN must be such an example, and speculate that another overluminous Ia remnant, SNR~0509-67.5, might show strong interaction with a massive CSM in the future.  It should be noted that, whereas Ia-CSM objects occupy only 0.1--1\% of SNe Ia \citep{Dilday2012}, the fraction of 91T-like events is $\sim$9\% \citep{Li2011a}.  Therefore, our result implies that roughly 10\% of SNe Ia are associated with a massive CSM.

Based on the observational clues obtained so far, we discuss possible progenitor systems for Kepler's SNR.  First of all, the presence of the CSM itself favors the SD scenario.  While some recent DD models predict the existence of CSMs \citep{Shen2013,Raskin2013,Tanikawa2015}, they originate from a H-rich envelope of a He WD or tidal tails from a C/O WD, hence they would not be CNO-processed N-rich material.  This is in conflict with the overabundance of N seen in the CSM of Kepler's SNR.  Moreover, the small amount of O ejecta, $<$0.069\,M$_\odot$ as shown in Table~\ref{tab:ejecta_mass}, is not expected for DD scenarios \citep[e.g.,][]{Pakmor2011,Roepke2012}, but is more consistent with delayed-detonation Chandrasekhar-mass models \citep[e.g.,][]{Maeda2010,Seitenzahl2013}.  Therefore, there is no observational support for DD channels as the origin of Kepler's SN, except for the absence of a candidate surviving companion star \citep{Kerzendorf2014}.

Outflows from SD progenitors can originate from either the exploding star or the donor star.  For Kepler's SN, the former case is less likely than the latter.  This is because the exploding star, which must have experienced a major mass loss $\sim3\times10^{4}$\,yrs before the explosion, cannot have had sufficient time to evolve to a WD and accrete material from its donor star to near the Chandrasekhar mass (cf., accretion timescale is of the order of 10$^{6}$\,yrs).  On the other hand, the exploding star is not necessarily a WD --- the so-called Type 1.5 model predicts that a degenerate C/O core of a massive ABG star with very low metallicity grows to near the Chandrasekhar mass before it loses its envelope \citep{Arnett1969,Iben1983,Lau2008}.  However, such a SN occurs inside a massive H envelope, so that the SN light curve should look like Type IIp SNe as shown in Fig.~\ref{fig:kepler_lc} (K.Maeda et al.\ in preparation).  Looking at Fig.~\ref{fig:kepler_lc}, this is evidently not the case for Kepler's SN.  Moreover, the fact that the metallicity of Kepler's CSM is nearly solar (Table~\ref{tab:csm}) also conflicts with the Type 1.5 model.

Turning to outflows from a companion/donor star, the high mass-loss rate we estimated, $\dot{M}\sim1.5\times10^{-5}$($v_\mathrm{w}$/10\,km\,s$^{-1}$)\,M$_\odot$\,yr$^{-1}$, suggests an AGB companion.  One possible progenitor system hosting an AGB companion is a recurrent nova \citep{Hachisu2001}.  However, there are two major observational facts contrary to the recurrent novae scenario.  First, Kepler's SN experienced a major mass eruption $\sim3\times10^{4}$\,yrs before the SN explosion.  Such a long nova timescale is not expected for recurrent nova systems with near Chandrasekhar-mass WDs, for which typical recurrence times are 10--100\,yrs.  Second, according to numerical simulations \citep{Marietta2000}, the impact of SN ejecta removes most of the envelope of a companion star if it is a giant like an AGB star.  As a result, the surviving companion will evolve away from the red giant branch, along a track of constant luminosity ($\sim10^{3}$\,$L_\odot$) for at least 10$^{5}$\,yrs.  Such a bright star, if present, should be easily detected by optical observations, but is not detected at least near the center of the remnant \citep{Kerzendorf2014}.  Thus, a recurrent nova is not likely the progenitor of Kepler's SN, either.  

There are other models that satisfy observational properties of Kepler's SNR, i.e., both high mass-loss rate and no bright surviving companion.  In efforts to explain the progenitor system of SNe Ia-CSM, \citet{Han2006} investigated the classical supersoft channel that consists of a C/O WD and a main sequence star.  They found that if the donor star is a relatively massive ($\sim$3\,M$_\odot$) main sequence star and the system experiences delayed dynamical instabilities, a large amount of mass can be lost from the system in the last few 10$^4$\,yrs before the explosion.  As a result, the donor star loses most of the mass before the explosion, and its luminosity becomes as low as the sun.  This model can explain both the massive CSM and no detectable surviving companion in Kepler's SNR.  Alternatively, \citet{Soker2013} argued that a WD and a hot core of a massive AGB star in a common envelope will violently and promptly merge if the WD is denser than the core.  In this case, there is no surviving companion left behind the SN while massive a CSM could be present near the progenitor system, making this model viable as well.  However, we have not yet found bservational evidence supporting these models, leaving us the mystery of the progenitor system.  Continuous searches for a surviving companion and/or signatures of a companion star in Kepler's SNR would be of great importance for revealing the progenitor of Kepler's SNR and SNe Ia in general.  

\section{Conclusions}

We have analyzed X-ray spectra and images of three young ($\sim$400\,yrs old), ejecta-dominated SNRs, i.e., Kepler's SNR, Tycho's SNR, and SNR~0509-67.5 in the LMC.  A simple spectral comparison among the three remnants suggests that Kepler's SNR and SNR~0509-67.5 produced substantially more IGE than Tycho's SNR.  In fact, we quantitatively revealed that the IGE masses for both Kepler's SNR and SNR~0509-67.5 are similar to those of overluminous SNe Ia, while that for Tycho's SNR is comparable to that of normal SNe Ia, assuming the total mass of IME and IGE to be $\sim$1.05\,M$_\odot$ \citep{Mazzali2007} and inferring the unshocked ejecta masses using the SN Ia's nucleosynthetic models together with the reverse shock positions estimated from {\it Chandra} images.  This inference is consistent with the fact that SNR~0509-67.5 and Tycho's SNR are confirmed remnants of overluminous and normal SNe Ia explosions, respectively.  Therefore, we conclude that Kepler's SN was likely an overluminous (91T-like) event.  Our method is quite simple, and thus could be a powerful tool for discriminating subclasses of Type Ia SNRs.

The {\it XMM-Newton}'s RGS spectrum shows a strong N Ly$\alpha$ line for Kepler's SNR.  We measured the N/H abundance to be $\sim$4 times the solar value, suggesting the presence of a CNO-processed CSM as was previously suggested by optical spectroscopy \citep{Blair1991}.  We revealed that the CSM is comprised of diffuse gas and dense knots.  The CSM masses are estimated to be $\sim$0.3\,M$_\odot$ and $\sim$0.035\,M$_\odot$ for the diffuse gas and dense knots, respectively.  The mass of the diffuse CSM allows us to infer a mass-loss rate of the progenitor system to be $\sim1.5\times10^{-5}$($v_\mathrm{w}$/10\,km\,s$^{-1}$)\,M$_\odot$\,yr$^{-1}$, consistent with those of AGB stars.  As for the knots, their slow optical proper motions and ionization timescales measured by using X-ray spectra suggest that they were located far away from the progenitor system at the time of SN explosion, and were recently shock heated.  We argue that Kepler's SN was likely a 91T-like event that started to interact with massive CSMs a few hundred years after the explosion.  This supports the possible link between 91T-like SNe and SNe Ia-CSM, implying that $\sim$10\% of SNe Ia are associated with massive CSM.  All of these SNe would be related to a single degenerate channel.

\acknowledgments

This work is supported by Japan Society for the Promotion of Science KAKENHI Grant Numbers 25800119 (S.Katsuda), 24740167 (K.Mori), 23740141, 26800100 (K.Maeda), 24740117, 15H02075, 15H00788 (M.Tanaka), 24540229 (K.Koyama), 23000004 (H.Tsunemi), 26670560 (H.Nakajima), 25105516 (Y.Maeda), and 24654052 (M.Ozaki).


\begin{deluxetable}{lcccc}
\tabletypesize{\small}
\tablecaption{X-ray data used in this paper}
\tablewidth{0pt}
\tablehead{
\colhead{Target}&\colhead{Instrument}&\colhead{Obs.~ID}&\colhead{Obs.~Date}&\colhead{Effective exposure (ks)}}
\startdata
Kepler's SNR & {\it XMM-Newton}'s RGS & 0084100101 & 2001-03-10 & 32.6 \\
Kepler's SNR & {\it Chandra}'s ACIS & 6715 & 2006-08-03 & 159.1 \\
Kepler's SNR & {\it Suzaku}'s XIS & 505092040 & 2011-02-28 & 146.2 \\
Tycho's SNR & {\it Suzaku}'s XIS & 500024010 & 2006-06-27 & 101.1 \\
SNR~0509-67.5 & {\it XMM-Newton}'s RGS & 0111130201 & 2000-07-04 & 36.0 \\
SNR~0509-67.5 & {\it Suzaku}'s XIS & 508072010 & 2013-04-11 & 175.9 \\
\enddata
\tablecomments{The PIs of the data are A.~Decourchelle (Kepler - XMM), S.~Reynolds (Kepler - Chandra), S.~Park (Kepler - Suzaku), Suzaku science working group (Tycho - Suzaku), M.~Watson (SNR~0509-67.5 - XMM), and H.~Yamaguchi (SNR~0509-67.5 - Suzaku).}
\label{tab:obs}
\end{deluxetable}

\begin{deluxetable}{lccccccccc}
\tabletypesize{\tiny}
\tablecaption{Spectral-fit parameters for the integrated X-ray spectra}
\tablewidth{0pt}
\tablehead{
\colhead{Parameter}& \colhead{Kepler's SNR} & \colhead{Kepler's SNR (south)} & \colhead{Tycho's SNR} & \colhead{SNR~0509-67.5}}
\startdata
CSM component & & & \\
~~~~$kT_{\mathrm e}$ (keV) & 1.06$\pm$0.03 & 0.48$^{+0.57}_{-0.20}$ & 0.41$\pm$0.01 & 1.76$^{+0.85}_{-0.48}$ \\
~~~~log($n_{\mathrm e}t$/cm$^{-3}\,$sec) & 10.81$^{+0.02}_{-0.01}$ & 10.65$^{+0.39}_{-0.17}$ & 10.69$^{+0.01}_{-0.05}$ & 10.22$^{+0.09}_{-0.34}$ \\
~~~~Abundance$^a$ (solar)~~~~~~~~N & 3.31$^{+0.24}_{-0.25}$ & 6.08$^{+3.17}_{-2.75}$ & 1$^b$ & 0.135$^b$ \\
~~~~Redshift (10$^{-3}$) & 1.39$\pm$0.05 & =Ejecta1 & =Ejecta1 & =Ejecta1 \\
~~~~Line broadening (E/1\,keV eV) & 2.91$^{+0.36}_{-0.32}$ & =Ejecta1 & =Ejecta1 & =Ejecta1 \\
~~~~$\int n_{\mathrm e} n_{\mathrm H} dV$/4$\pi d^{2}$ ($10^{10}$cm$^{-5}$) & 193.15$^{+4.03}_{-6.48}$ & 9.37$^{+11.11}_{-2.11}$ & 1788.32$^{+74.61}_{-12.55}$ & 1.55$^{+0.14}_{-0.43}$ \\
\hline 
Ejecta components & & & \\
1~~~$kT_{\mathrm e}$ (keV) & 0.37$\pm$0.01 & 0.44$^{+0.01}_{-0.02}$ & 0.70$^{+0.02}_{-0.01}$ & 1.09$^{+0.11}_{-0.24}$ \\
~~~~log($n_{\mathrm e}t$/cm$^{-3}\,$sec) & 10.52$\pm$0.01 & 10.39$^{+0.06}_{-0.03}$  & 10.79$\pm$0.01  & 10.14$^{+0.02}_{-0.01}$ \\
~~~~Abundance (10$^{4}$ solar)~~~~C & 0.54$^b$ & 0.54$^b$ & 0.54$^b$ & 0.54$^b$ \\
~~~~~~~~~~~~~~~~~~~~~~~~~~~~~~~~~~~~~N & {0$^b$}& {0$^b$}& {0$^b$}& {0$^b$} \\
~~~~~~~~~~~~~~~~~~~~~~~~~~~~~~~~~~~~~O & 0.25$^{+0.02}_{-0.03}$ & 0.31$^{+0.12}_{-0.04}$ & 0.75$\pm$0.01 & 0.22$^{+0.3}_{-0.15}$ \\
~~~~~~~~~~~~~~~~~~~~~~~~~~~~~~~~~~~~~Ne& 0.67$^{+0.06}_{-0.07}$ & 0.84$^{+0.14}_{-0.16}$ & 0$(<0.0003)$ & 1.15$^{+0.18}_{-0.17}$ \\
~~~~~~~~~~~~~~~~~~~~~~~~~~~~~~~~~~~~~Mg& 0.77$^{+0.07}_{-0.06}$ & 0.53$\pm$0.14 & 0.34$\pm$0.01 & 0.07$^{+0.30}_{-0.07}$ \\
~~~~~~~~~~~~~~~~~~~~~~~~~~~~~~~~~~~~~Si& 10$^b$& 10$^b$& 10$^b$& 10$^b$  \\
~~~~~~~~~~~~~~~~~~~~~~~~~~~~~~~~~~~~~S& 15.53$^{+0.22}_{-0.14}$ & 12.31$^{+0.64}_{-0.66}$ & 10.65$^{+0.02}_{-0.03}$ & 15.79$^{+1.23}_{-0.94}$ \\
~~~~~~~~~~~~~~~~~~~~~~~~~~~~~~~~~~~~~Ar& 18.98$^{+0.46}_{-0.52}$ & 11.34$^{+1.69}_{-1.76}$ & 11.02$^{+0.07}_{-0.15}$ & 10.59$^{+2.16}_{-1.96}$ \\
~~~~~~~~~~~~~~~~~~~~~~~~~~~~~~~~~~~~~Ca& 36.40$^{+1.51}_{-1.60}$ & 23.74$^{+5.74}_{-5.62}$ & 25.83$^{+0.35}_{-0.50}$ & 14.59$^{+4.8}_{-2.46}$ \\
~~~~~~~~~~~~~~~~~~~~~~~~~~~~~~~~~~~~~Fe& 13.65$^{+0.17}_{-0.47}$ & 6.63$^{+2.58}_{-1.87}$ & 1.49$\pm$0.01 & 175.56$^{+1924.02}_{-56.18}$ \\
~~~~$\int n_{\mathrm e} n_{\mathrm H} dV$/4$\pi d^{2}$ ($10^{5}$cm$^{-5}$) & 961.42$^{+114.02}_{-9.8}$ & 230.29$^{+104.58}_{-62.48}$ & 9014.62$^{+45.8}_{-775.91}$ & 0.13$^{+0.02}_{-0.04}$ \\
~~~~Redshift (10$^{-3}$) & -2.94$\pm$0.01 & -5.60$^{+1.84}_{-5.91}$ & -4.57$\pm$0.01 & -0.89$^{+0.23}_{-0.41}$ \\
~~~~Line broadening (E/1\,keV eV) & 9.13$^{+0.14}_{-0.21}$ & 10.16$^{+0.48}_{-0.45}$ & 9.90$^{+0.05}_{-0.03}$ & 11.90$^{+0.67}_{-0.78}$ \\
2~~~$kT_{\mathrm e}$ (keV) & 2.08$^{+0.01}_{-0.02}$ & 1.46$^{+0.33}_{-0.08}$ & 0.96$\pm$0.01 & 5.44$^{+0.82}_{-2.26}$ \\
~~~~log($n_{\mathrm e}t$/cm$^{-3}\,$sec) & 10.32$\pm$0.01 & =Ejecta1  & 10.90$\pm$0.01  & =Ejecta1 \\
~~~~Abundance$^c$ (10$^{4}$ solar)~~~Fe & 3.58$^{+0.04}_{-0.05}$ & 2.23$^{+0.15}_{-0.18}$ & 0$(<0.0002)$ & 0$(<0.03)$ \\
~~~~$\int n_{\mathrm e} n_{\mathrm H} dV$/4$\pi d^{2}$ ($10^{5}$cm$^{-5}$) & 875.35$^{+4.73}_{-4.19}$ & 131.51$^{+21.9}_{-8.78}$ & 22.75$\pm$0.02 & 0.013$\pm$0.001 \\
~~~~Redshift (10$^{-3}$) & =Ejecta1 & =Ejecta1 & =Ejecta1 & =Ejecta1 \\
~~~~Line broadening (E/1\,keV eV) & =Ejecta1 & =Ejecta1 & =Ejecta1 & =Ejecta1 \\
3~~~$kT_{\mathrm e}$ (keV) & 2.59$\pm$0.01 & 1.92$\pm$0.1 & 9.34$^{+0.03}_{-0.25}$ & 6.97$^{+1.7}_{-1.08}$ \\
~~~~log($n_{\mathrm e}t$/cm$^{-3}\,$sec) & 9.21$^{+0}_{-0}$ & 9.02$^{+0.04}_{-0.03}$  & 9.80$\pm$0.01  & 9.60$^{+0.05}_{-0.02}$ \\
~~~~Abundance$^d$ (10$^{4}$ solar)~~Ar & 0 ($<0.19$) & 0.45$^{+0.46}_{-0.43}$ & 0 ($<2.7$) & 37.61$^{+1.40}_{-1.09}$ \\
~~~~~~~~~~~~~~~~~~~~~~~~~~~~~~~~~~~~~Ca & 1.43$^{+0.31}_{-0.28}$ & 1.65$^{+0.59}_{-0.58}$ & 64.82$^{+1.69}_{-1.75}$ & 47.28$^{+16.22}_{-14.14}$ \\
~~~~~~~~~~~~~~~~~~~~~~~~~~~~~~~~~~~~~Fe & 10$^b$ & 10$^b$ & 10$^b$ & 10$^b$ \\
~~~~Redshift (10$^{-3}$) & -5.73$^{+0.14}_{-0.18}$ & -9.50$^{+0.98}_{-1.09}$ & -5.24$^{+0.06}_{-0.05}$ & 0.32$^{+2.72}_{-1.08}$ \\
~~~~Line broadening (E/1\,keV eV) & 12.10$\pm$0.27 & 11.15$^{+1.31}_{-1.42}$ & 10.76$^{+0.23}_{-0.24}$ & 16.81$^{+1.41}_{-1.34}$ \\
~~~~$\int n_{\mathrm e} n_{\mathrm H} dV$/4$\pi d^{2}$ ($10^{5}$cm$^{-5}$) & 5808.47$^{+71.11}_{-63.17}$ & 1748.80$^{+137.58}_{-561.66}$ & 1.08$\pm$0.01 & 0.006$^{+0.002}_{-0.001}$ \\
\hline 
Power-law component & & & \\
~~~~$\Gamma$ & 2.64$^{+0.02}_{-0.01}$ & 2.55$^{+0.02}_{-0.01}$ & 2.56$\pm$0.01 & 3.37$^{+0.16}_{-0.2}$ \\
~~~~Norm (ph\,keV$^{-1}$\,cm$^{-2}$\,s$^{-1}$ at 1\,keV) & 98.93$^{+5.51}_{-1.97}$ & 24.49$^{+0.43}_{-0.36}$ & 714.68$^{+1.96}_{-1.75}$ & 1.85$^{+0.37}_{-0.48}$  \\
\hline 
Additional lines & & & \\
~~~~Fe~L+O~K~~~Center (keV) & 0.708$^{+0.002}_{-0.001}$ & --- & 0.741$\pm$0.001 & 0.724$\pm$0.002 \\
~~~~~~~~~~~~~~~~~~~~Norm (10$^{-4}$ ph\,keV$^{-1}$\,cm$^{-2}$\,s$^{-1}$) & 68.18$^{+5.4}_{-5.0}$ & --- & 986.88$\pm$19.1  & 2.38$\pm$0.3  \\
~~~~Fe~L+Ne~K~~Center (keV) & --- & 1.201$\pm$0.011 & 1.194$^{+0.001}_{-0.003}$ & --- \\
~~~~~~~~~~~~~~~~~~~~Norm (10$^{-4}$ ph\,keV$^{-1}$\,cm$^{-2}$\,s$^{-1}$) & --- & 1.96$^{+0.5}_{-0.4}$ & 33.89$\pm$0.6  & --- \\
~~~~Fe~L+Ne~K~~Center (keV) & 1.227$^{+0.001}_{-0.003}$ & --- & 1.254$\pm$0.001 & 1.244$\pm$0.001 \\
~~~~~~~~~~~~~~~~~~~~Norm (10$^{-4}$ ph\,keV$^{-1}$\,cm$^{-2}$\,s$^{-1}$) & 22.75$^{+1.2}_{-1}$ & --- & 42.40$\pm$0.5  & 0.19$\pm$0.1  \\
~~~~Cr~K~~~~~~~~~~Center (keV) & 5.514$^{+0.035}_{-0.037}$ & 5.438$^{+0.073}_{-0.08}$ & 5.474$^{+0.016}_{-0.019}$ & 5.436$^{+0.068}_{-0.071}$ \\
~~~~~~~~~~~~~~~~~~~~Norm (10$^{-7}$ ph\,keV$^{-1}$\,cm$^{-2}$\,s$^{-1}$) & 85.18$^{+16.7}_{-17.1}$ & 17.96$\pm$10.2 & 197.98  & 7.05$\pm$3.2  \\
~~~~Mn~K~~~~~~~~~Center (keV) & 5.976$^{+0.04}_{-0.042}$ & 5.862$^{+0.266}_{-0.152}$ & 5.935$^{+0.07}_{-0.071}$ & 5.893$^{+0.068}_{-0.071}$ \\
~~~~~~~~~~~~~~~~~~~~Norm (10$^{-7}$ ph\,keV$^{-1}$\,cm$^{-2}$\,s$^{-1}$) & 60.18$\pm$16.3 & 14.16$^{+10.8}_{-14.2}$ & 41.07$\pm$32  & 3.56$\pm$3.4  \\
\hline 
$\chi^{2}$/d.o.f. & 8289.8 / 5724 & 1744.2 / 1305 & 5659.0 / 1981 & 1031.6 / 783 \\
\enddata
\tablecomments{$N_{\mathrm H}$ values are fixed to 6.4$\times10^{21}$\,cm$^{-2}$, 1$\times10^{22}$\,cm$^{-2}$, and 0.6$\times10^{21}$\,cm$^{-2}$ for Kepler's SNR, Tycho's SNR, and SNR~0509-67.5, respectively.  $^a$Other elements are fixed to either the solar values \citep{Wilms2000} or typical ISM values of the Large Magellanic Clouds \citep{Korn2002,Russell1992} for the two Galactic SNRs or SNR~0509-67.5, respectively.  $^b$Fixed values.  $^c$Other elemental abundances are lined to those in the Ejecta1 component.  $^d$Other elemental abundances are fixed to zero.}
\label{tab:param}
\end{deluxetable}


\begin{deluxetable}{lccccc}
\tabletypesize{\small}
\tablecaption{Ejecta masses (M$_\odot$)}
\tablewidth{0pt}
\tablehead{
\colhead{Element}&\colhead{Kepler's SNR}&\colhead{Tycho's SNR}&\colhead{SNR~0509-67.5}}
\startdata
O & 0.015$^{+0.054}_{-0.008}$ & 0.319$^{+0.012}_{-0.159}$ & 0.034$^{+0.001}_{-0.018}$ \\
Ne & 0.007$^{+0.011}_{-0.004}$ & 0$(<0.001)$ & 0.033$^{+0.002}_{-0.023}$ \\
Mg & 0.002$(<0.005)$  & 0.008$^{+0.001}_{-0.005}$ & 0.001$(<0.002)$ \\
Si & 0.040$^{+0.060}_{-0.014}$ & 0.284$^{+0.053}_{-0.123}$ & 0.106$^{+0.026}_{-0.072}$\\
S & 0.047$^{+0.072}_{-0.016}$ & 0.229$^{+0.042}_{-0.096}$ & 0.126$^{+0.031}_{-0.090}$ \\
Ar & 0.012$^{+0.022}_{-0.006}$ & 0.072$^{+0.009}_{-0.024}$ & 0.018$^{+0.006}_{-0.011}$ \\
Ca & 0.021$^{+0.031}_{-0.012}$ & 0.118$^{+0.012}_{-0.041}$ & 0.089$^{+0.021}_{-0.072}$\\
Fe (shocked) & 0.710$^{+0.094}_{-0.249}$ & 0.134$^{+0.387}_{-0.028}$ & 0.372$^{+0.355}_{-0.118}$\\
Ni$^a$ (shocked) & 0.110$^{+0.015}_{-0.038}$ & 0.021$^{+0.060}_{-0.005}$ & 0.058$^{+0.055}_{-0.019}$ \\
IGE$^b$ (unshocked) & 0.129$^{+0.234}_{-0.087}$ & 0.194$^{+0.099}_{-0.120}$ & 0.318$^{+0.095}_{-0.250}$
\enddata
\tablecomments{$^a$The Ni/Fe abundance ratio is assumed to be that expected in the W7 model \citep{Nomoto1984}.  $^b$Taken from the cold IGE panels in Fig.~\ref{fig:mass_vs_dist}.  
}
\label{tab:ejecta_mass}
\end{deluxetable}


\begin{deluxetable}{lcccccc}
\tabletypesize{\tiny}
\tablecaption{Spectral-fit parameters for the CSM knots in Kepler's SNR}
\tablewidth{0pt}
\tablehead{
\colhead{Parameter}&\colhead{Knot-1}&\colhead{Knot-2}&\colhead{Knot-3}&\colhead{Knot-4}&\colhead{Knot-5}}
\startdata
$kT_{\mathrm e}$ (keV) & 1.10$\pm$0.11 & 0.95$^{+0.1}_{-0.08}$ & 1.92$^{+0.57}_{-0.38}$ & 0.68$^{+0.13}_{-0.09}$ & 1.84$^{+0.97}_{-0.52}$ \\
log($n_{\mathrm e}t$/cm$^{-3}\,$sec) & 10.59$^{+0.07}_{-0.09}$ & 10.96$^{+0.13}_{-0.12}$ & 10.66$^{+0.09}_{-0.12}$ & 10.91$^{+0.19}_{-0.17}$  & 10.36$^{+0.2}_{-0.18}$  \\
Abundance (solar)\,~~~N & 3.5$^a$ & 3.5$^a$ & 3.5$^a$ & 3.5$^a$ & 3.5$^a$ \\
~~~~~~~~~~~~~~~~~~~~~~~~~~~O & 0.97$^{+0.15}_{-0.13}$ & 0.95$^{+0.21}_{-0.08}$ & 1 & 1 & 1 \\
~~~~~~~~~~~~~~~~~~~~~~~~~~~Ne & 1.31$^{+0.23}_{-0.22}$ & 0.84$^{+0.22}_{-0.21}$ & 1 & 1 & 1 \\
~~~~~~~~~~~~~~~~~~~~~~~~~~~Mg & 1.43$^{+0.2}_{-0.18}$ & 0.91$^{+0.16}_{-0.15}$ & 1 & 1 & 1 \\
$\int n_{\mathrm e} n_{\mathrm H} dV$/4$\pi d^{2}$ ($10^{10}$cm$^{-5}$) & 1.87$^{+0.17}_{-0.15}$ & 1.58$\pm$0.13 & 0.55$^{+0.06}_{-0.05}$ & 0.77$^{+0.14}_{-0.12}$ & 0.26$^{+0.03}_{-0.02}$ \\
$n_{\mathrm e}$ (cm$^{-3}$) & 23.4 & 29.2 & 6.4 & 11.3 & 6.9 \\
Mass ($10^{-4}\,M_{\odot}$)& 2.8 & 1.9 & 3.1 & 2.4 & 1.3  \\
\hline 
$\chi^{2}$/d.o.f. & 181.8 / 134 & 163.1 / 111 & 130.1 / 128 & 95.9 / 109 & 95.6 / 101  \\
\enddata
\tablecomments{$N_{\mathrm H}$ values are fixed to 6.4$\times10^{21}$\,cm$^{-2}$.  We assume $n_{\mathrm e} = 1.2 n_{\mathrm H}$ and the distance $d = 4.2$\,kpc.  $^a$Fixed values.}
\label{tab:csm}
\end{deluxetable}

\end{document}